\UseRawInputEncoding
\documentclass[useAMS,usenatbib]{mnras}

\usepackage{graphicx}	
\usepackage{amsmath}	
\usepackage{amssymb}	
\usepackage[dvipsnames]{xcolor}
\usepackage{makecell}

\newcommand{\be}{\begin{equation}}
\newcommand{\ee}{\end{equation}}
\newcommand{\bea}{\begin{eqnarray}}
\newcommand{\eea}{\end{eqnarray}}

\def\aap{A\&A}
\def\apj{ApJ}

\def\apjl{ApJ}
\def\mnras{MNRAS}

\def\apjs{ApJS}
\def\prd{Phys. Rev. D}

\renewcommand{\vec}[1]{ {\bmath #1} } 

\def\ltsima{$\; \buildrel < \over \sim \;$}
\def\simlt{\lower.5ex\hbox{\ltsima}}
\def\gtsima{$\; \buildrel > \over \sim \;$}
\def\simgt{\lower.5ex\hbox{\gtsima}}

\title[Smulations of Constrained Interacting Dark Energy]{The CIDER simulations: Nonlinear structure formation in the Constrained Interacting Dark Energy Scenario}
\author[M. Baldi]{\parbox{\textwidth}{Marco Baldi$^{1,2,3}$\thanks{marco.baldi5@unibo.it}}\\
\\
$^{1}$Dipartimento di Fisica e Astronomia, Alma Mater Studiorum - University of Bologna, Via Piero Gobetti 93/2, 40129 Bologna BO, Italy\\
$^{2}$INAF - Osservatorio Astronomico di Bologna, Via Piero Gobetti 93/3, 40129 Bologna BO, Italy\\
$^{3}$INFN - Istituto Nazionale di Fisica Nucleare, Sezione di Bologna, Viale Berti Pichat 6/2, 40127 Bologna BO, Italy}

\begin{document}
\pagerange{\pageref{firstpage}--\pageref{lastpage}} \pubyear{2011}
\maketitle
\label{firstpage}
\begin{abstract}	
\noindent 
We present for the first time a suite of cosmological simulations for a particular class of interacting Dark Energy cosmologies characterised by a background expansion history constrained to be indistinguishable from $\Lambda $CDM. Such {\em Constrained Interacting Dark Energy} scenario -- or CIDER -- has been recently proposed by \citet{Barros_etal_2019} and has the appealing feature of suppressing structure formation at late times, thereby possibly alleviating the persisting $\sigma _{8}$ tension while leaving background observables unaffected. A crucial step to assess the viability of such scenarios is then represented by quantifying their impact on structure formation at non-linear scales, which is what we start investigating with the simulations discussed in the present work. We show that -- for reasonable parameter choices -- the reconstructed scalar potential is close to an exponential for most of the matter dominated epoch, and that the nonlinear evolution of structures in these models imprints specific footprints on matter and halo statistics that may allow to break degeneracies with standard cosmological parameters. 
\end{abstract}

\begin{keywords}
dark energy -- dark matter --  cosmology: theory -- galaxies: formation
\end{keywords}


\section{Introduction}
\label{i}

Over the past decades cosmology has reached its maturity thanks to the coordinated developments of robust theoretical predictions and ambitious observational campaigns that provided us with an ever increasing wealth of high-quality data capable to rule out different competing models and to establish a standard ``concordance" cosmological scenario. The $\Lambda $CDM
 model -- based on a Cosmological Constant $\Lambda $ as the source of the observed accelerated expansion and on Cold Dark Matter (CDM) particles as the dominating matter component in the universe, driving the growth of cosmic structures --  has emerged as the simplest description of the vast majority of presently available data, and the next generation of observational endeavours -- such as e.g. Euclid \citep[][]{Laureijs_etal_2011}, The Vera Rubin Observatory \citep[][]{LSST}, the Square Kilometre Aarray \citep[][]{SKA} -- hold the promise to constrain its parameters with percent accuracy, opening the so-called era of {\em ``precision cosmology"}.

Nonetheless, despite the remarkable success of the $\Lambda $CDM cosmology, its foundations still pose several puzzling questions. On the one side, the cosmological constant $\Lambda $ does not seem to fit naturally in the framework of particle physics, with its observed value -- deemed to remain constant throughout the whole evolution of the Universe -- being exceedingly small with respect to the typical energy scale of fundamental particles and fields in the early Universe \citep[][]{Weinberg_1989,Sahni_2002}. On the other side, the hypothetical CDM particles have so far been evading all experimental efforts for a direct \citep[see e.g.][]{MarrodanUndagoitia_Rauch_2016,Buchmueller_Doglioni_Wang_2017} or indirect \citep[see e.g.][]{Gaskins_2016} detection, casting doubts on the actual existence of such fundamental particles and reviving the interest for alternative explanations \citep[as e.g. for the recent re-consideration of  Primordial Black Holes as possible Dark Matter candidates, see ][]{Green_Kavanagh_2021}.

Besides being affected by these fundamental puzzles, the $\Lambda $CDM scenario has been recently questioned also for the persisting tensions between its best-fit parameters as derived independently from high-redshift and low-redshift observational probes \citep[see e.g. the extensive review given in][]{SNOWMASS_tensions_2022}. In fact, several independent studies have consistently shown how the former -- mostly based on Cosmic Microwave Background (CMB) observations \citep[][]{Planck_2018_VI,ACT_DR4_2020,SPT-3G_2021} -- result in a lower value of the Hubble constant $H_{0}$ and in a higher value of the scalar perturbations amplitude parameter $\sigma _{8}$ with respect to the latter, which include different local probes of the cosmic expansion (such as Supernovae Ia, e.g. \citet{SHOES_2021}, cosmic chronometers, e.g. \citet{Moresco_etal_2022}, gravitational waves, e.g. \citet{LIGO_H0_2021}, strong lensing time delay, e.g. \citet{Wong_etal_2020}) and of the Large-Scale Structure (LSS) evolution (including Weak Gravitational Lensing, e.g. \citet{Heymans_etal_2021,Hildebrandt_etal_2017,Hildebrandt_etal_2020,Joudaki_etal_2017,Joudaki_etal_2018,Troxel_etal_2018,KiDS-1000,DES_Y3_Amon,DES_Y3_Secco}, galaxy clustering, e.g. \citet{Abbott_etal_2018,Chen_etal_2022, Troster_etal_2020}, clusters of galaxies, e.g. \citet{Vikhlinin_etal_2009,Abbott_etal_2020,Lesci_etal_2022, Marulli_etal_2021, SPT-SZ, DES_Y1_Clusters}, and redshift-space distortions, see e.g. \citet{Macaulay_etal_2013,Kazantzidis_Perivolaropoulos_2018}). Although such tensions are still not significant enough to claim for a failure of the $\Lambda $CDM model, and  could very well be due to unaccounted systematics in some of the involved observational pipelines, they have been persisting over the last few years and still remain unaddressed, thereby motivating the exploration of alternative models that may be capable of providing a better fit to the data.

These include a wide variety of possible extensions to the minimal set of hypotheses and assumptions that constitute the foundations of the $\Lambda $CDM cosmology, ranging from alternative Dark Energy models \citep[such as Quintessence or k-essence, see e.g.][]{Wetterich_1988,Ratra_Peebles_1988,kessence} characterised by a time evolution of the energy density -- and of the corresponding negative pressure -- associated with the acceleration of the cosmic expansion, to extensions of the theory of gravity beyond standard Einstein's General Relativity, giving rise to a modified evolution of the cosmic geometry and of the gravitational growth of cosmic structures at large scales \citep[see e.g. ][for a general overview]{Sotiriou_Faraoni_2010}, to Dark Matter scenarios featuring ultra-light bosonic particles \citep[like the Axions][]{Hu_Barkana_Gruzinov_2000,Hui_etal_2017} or mixtures of different families of fundamental particles \citep[such as standard and sterlie neutrinos, see e.g.][]{Boyarsky_Ruchayskiy_Shaposhnikov_2009} to the more radical paradigm shift involving Primordial Black Holes as the source of the Dark Matter phenomenology \citep[][]{Green_Kavanagh_2021,Carr_Kuhnel_Sandstad_2016,Bird_etal_2016}. 

Among such alternative scenarios, a significant interest has been raised by interacting Dark Energy models \citep[][]{Wetterich_1995,Amendola_2000,Farrar2004,Baldi_2011a,Simpson_2010,Skordis_Pourtsidou_Copeland_2015} characterised by various possible forms of energy-momentum exchange between a light scalar field playing the role of Dark Energy and one \citep[or multiple, as e.g. in ][]{Huey_Wandelt_2006,Amendola_Baldi_Wetterich_2008,Baldi_2012a,Amendola_Barreiro_Nunes_2014} species  of massive particles playing the role of Dark Matter. As a consequence of the interaction, the energy density and equation of state of the Dark Energy field may exhibit a non-trivial dynamical evolution and the Dark Matter particles may experience additional forces giving rise to a modified growth of structures both in the linear and non-linear regimes \citep[][]{Amendola_2004,Pettorino_Baccigalupi_2008,Nusser_Gubser_Peebles_2005,Li_Zhao_2009,Baldi_etal_2010,Li_Barrow_2010b,Penzo_etal_2015,Tarrant_etal_2012,Sutter_etal_2015}. Both these types of effects -- that are proportional to the coupling strength between the two dark sectors -- can be observationally tested, making this kind of models predictive and falsifiable, and have led to place tight constraints \citep[][]{Pettorino_etal_2012,Pettorino_2013,Planck_2015_XIV,Gomez-Valent_Pettorino_Amendola_2020} on the simplest realisations of this scenario  where the coupling is assumed to be a constant and the scalar field is equipped with a self-interaction potential given by some simple analytic function, as e.g. an exponential \citep[][]{Lucchin_Matarrese_1984,Wetterich_1988} or an inverse power \citep[][]{Ratra_Peebles_1988}. Nonetheless, both these assumptions could be released, resulting in weaker observational constraints and a richer phenomenology of the models.

In particular, in this work we will consider the situation in which the latter assumption is dropped, while we will explore a theoretically motivated example of a non-constant coupling function in an upcoming companion paper (Baldi et al. {\em in prep}).
More specifically, in a recent work \citet{Barros_etal_2019} have proposed a particular type of such interacting Dark Energy models where the scalar self-interaction potential is not specified {\em a priori} but is derived from the dynamical evolution of the model after fixing the background expansion history to that of $\Lambda $CDM. This implies that the model becomes indistinguishable from $\Lambda $CDM at the background level (by construction), thereby evading all constraints derived from geometric probes, still retaining its effects on the growth of density perturbations and cosmic structures. Interestingly, this approach leads to a suppression of structure growth, contrary to what predicted by most of the standard interacting Dark Energy scenarios (as well as by most modified gravity theories), thereby providing a possible handle on the $\sigma _{8}$ tension mentioned above.

In the present work, we present for the first time a set of cosmological N-body simulations that incorporate all the effects characterising the particular class of interacting Dark Energy models proposed in \citet{Barros_etal_2019}, and we discuss the effects imprinted by the interaction on a range of basic cosmological observables such as the matter power spectrum, the halo mass function and concentrations, the halo and cosmic voids density profiles, and the voids abundance. We also numerically derive the shape of the resulting scalar self-interaction potential, showing that it is well approximated by an exponential function with a slowly varying slope. Our results show that this class of models could account for a lower value of $\sigma _{8}$ and in general for a slower evolution of cosmic structure at low redshifts, still remaining consistent with $\Lambda $CDM at the background level, without requiring (or resulting in) contrived or unnatural shapes of the scalar field potential, thereby offering a viable and appealing alternative to the standard cosmological scenario.\\

The paper is organised as follows. In Section~\ref{sec:models} we briefly introduce the cosmological models under investigation, highlighting how they are indistinguishable by construction from $\Lambda $CDM at the background level while showing deviations in the evolution of perturbations. In Section~\ref{sec:sims} we describe the numerical setup of the cosmological simulations that we have performed for such models. In Section~\ref{sec:results} we describe the main results extracted from the simulations, discussing how the coupling affects some basic cosmological observables such as the matter power spectrum, the baryon-CDM bias, the halo mass function, the profiles and concentrations of halos, as well as the abundance and density profiles of cosmic voids. Finally, in Section~\ref{sec:conclusions} we summarise our findings and draw our conclusions.

\section{The {\em CiDEr} cosmological scenario}
\label{sec:models}
We introduce in this Section the main features of the cosmological models under investigation in the present work, thereby summarising the general description of the background (Section~\ref{bkd}) and linear perturbations (Section~\ref{sec:linear}) evolution of these scenarios first discussed in \citet{Barros_etal_2019}, to which we refer for a more detailed derivation of the main equations. Despite the variety of conventions and notations that have been adopted in the past by various authors (including ourselves) for the description of these models, we deliberately choose to follow the notation used in \citet{Barros_etal_2019} to allow for an easier cross-reference of our results.

\subsection{Background evolution}
\label{bkd}
We consider a standard Coupled Quintessence cosmological model \citep[see e.g.][for a general discussion on Coupled Quintessence]{Amendola_2000,Amendola_2004,Farrar2004,Baldi_etal_2010,Baldi_2011a} for a universe described by a flat Friedmann-Lema\^itre-Robertson-Walker metric:
\begin{equation}
\label{flrw}
ds^{2} = -dt^2 + a(t) \delta _{ij}dx^{i}dx^{j}
\end{equation}
where $a(t)$ is the cosmic scale factor and its derivative with respect to the cosmic time $t$ defines the Hubble function $H(a)\equiv a^{-1}d{a}/dt = \dot{a}/a$.
The energy-density of the universe is described by the total stress-energy tensor $T_{\mu \nu }=\sum _{i} T^{(i)}_{\mu \nu}$ where $i$ is an index that runs over the different cosmic components which can be individually described as perfect-fluids:
\begin{equation}
\label{tmunu}
    T^{(i)}_{\mu \nu} = (\rho _{i} + p_{i})u^{(i)}_{\mu}u^{(i)}_{\nu} - p_{i}g_{\mu \nu}\,.
\end{equation}
Here $\rho _{i}$ and $p_{i}$ are the energy and pressure density of the fluid, respectively, $u^{(i)}_{\mu}$ is its four-velocity, and $g_{\mu \nu}$ is the metric tensor.
In particular, we consider four different components of the total energy budget, namely radiation and other relativistic species ($r$), baryonic particles ($b$), Cold Dark Matter ($c$) and Dark Energy in the form of a classical light scalar field ($\phi $).
Following the standard approach of Coupled Quintessence models we assume that relativistic species and baryons are minimally coupled and therefore separately conserved
\begin{equation}
    \nabla _{\mu}T^{(r)\mu }_{\nu} = \nabla _{\mu }T^{(b)\mu }_{\nu} = 0
\end{equation}
while Dark Energy (DE) and Cold Dark Matter (CDM) are allowed to directly interact through an exchange of energy-momentum which is bound to keep the total stress-energy tensor $T_{\mu \nu}$ conserved as well, which means:
\begin{equation}
    \nabla _{\mu }T^{(c)\mu }_{\nu} = C_{\nu}(\phi ) = - \nabla _{\mu }T^{(\phi )\mu }_{\nu} \,.
\end{equation}
where $C_{\nu }(\phi )$ is a conformal coupling function which can be expressed in the form \citep[see again][]{Amendola_2000,Baldi_2011a}:
\begin{equation}
    C_{\nu }(\phi ) = \kappa \beta (\phi )\rho _{c}\nabla _{\nu }\phi \,.
\end{equation}
In the previous expression, we have defined $\kappa \equiv 8\pi G$, with $G$ the Newton's gravitational constant, and $\beta (\phi )$ the coupling function that we will assume to be a constant $\beta (\phi )=\rm{const}$ in this work.

Restricting the coupling to the dark sector, thereby leaving baryons uncoupled, ensures consistency with solar-system tests of gravity \citep[see e.g.][]{Bertotti_Iess_Tortora_2003,Adelberger_etal_2003,Will_2014} without the need of a screening mechanism that is instead required in Modified Gravity scenarios. At the same time, having different matter species interacting with different couplings determines an effective violation of the Weak Equivalence Principle that can result in a rich phenomenology potentially allowing to constrain the model \citep[see e.g.][]{Amendola_Tocchini-Valentini_2002,Kesden_Kamionkowski_2006,Keselman_Nusser_Peebles_2009,Keselman_Nusser_Peebles_2010,Archidiacono_etal_2022}. 

The DE scalar field $\phi $ has an intrinsic energy density and pressure given by
\begin{eqnarray}
\label{phidensity}
    \rho _{\phi } = \frac{1}{2}g^{\mu \nu }\partial _{\mu }\phi \partial _{\nu }\phi + V(\phi )\\
    \label{phipressure}
    p_{\phi } = \frac{1}{2}g^{\mu \nu }\partial _{\mu }\phi \partial _{\nu }\phi - V(\phi )
\end{eqnarray}
where $V(\phi )$ is a self-interaction potential. The interaction terms appearing in the conservation equations of the different species then result in the following set of background dynamic equations
\begin{eqnarray}
\label{kleingordon}
\ddot{\phi } + 3H\dot{\phi } + \frac{dV}{d\phi }&=&\kappa \beta \rho _{c} \\
\label{continuity}
\dot{\rho }_{c} + 3H\rho _{c}&=&-\kappa \beta \rho _{c} \\
\label{continuity_b}
\dot{\rho }_{b} + 3H\rho _{b}&=&0 \\
\label{continuity_r}
\dot{\rho }_{r} + 4H\rho _{r}&=&0 \,.
\end{eqnarray}
These are the standard background evolution equations of Coupled Dark Energy models and their solutions have been investigated in a large number of works over the past two decades \citep[as e.g.][just to mention some]{Amendola_2000,Amendola_2004,Farrar2004,Mainini_Bonometto_2004,Pettorino_Baccigalupi_2008,Baldi_etal_2010,Baldi_2011a,Skordis_Pourtsidou_Copeland_2015}. 
In particular, the integration of Eq.~\ref{continuity} leads to an evolution for the CDM density given by
\begin{equation}
\label{cdmevolution}
    \rho _{c}(a) = \rho _{c,0}a^{-3}\exp{(-\kappa \beta \phi)}
\end{equation}
where the last exponential factor is a consequence of the energy exchange with the DE field. The immediate consequence of Eq.~(\ref{cdmevolution}) for a species that conserves particle number -- as CDM -- is that the rest mass of individual particles changes in time as $m_{c}(a) = m_{c\,,0}\exp{(-\kappa \beta \phi)}$.\\

In the standard approach, this kind of models  are defined by choosing a theoretically-motivated analytical form for the self-interaction potential function $V(\phi )$ -- as for instance the inverse-power law \citep[][]{Ratra_Peebles_1988}, the exponential \citep[][]{Lucchin_Matarrese_1984,Wetterich_1995}, or the SUGRA \citep[][]{Brax_Martin_1999} potentials -- and then derive the cosmological evolution following from such motivated (yet somewhat still arbitrary) choices. In particular, the aforementioned analytic potentials have been found to
provide natural scaling solutions for the evolution of the different cosmic species \citep[see e.g.][]{Copeland_etal_1998,Ferreira_Joyce_1998,Amendola_2000,Baccigalupi_Matarrese_Perrotta_2000}, and other appealing phenomenological behaviours \citep[see e.g.][]{Baldi_2011c} while at the same time determining a sizeable impact on the background expansion history which led to tightly constrain the allowed range of coupling values -- at least for the constant coupling case -- to a level which may be hardly distinguishable from the standard uncoupled case also for what concerns linear and non-linear structure formation \citep[][]{Amendola_Quercellini_2003,Pettorino_etal_2012,Pettorino_etal_2013,Planck_2015_XIV,Gomez-Valent_Pettorino_Amendola_2020}.

In the present work, we will follow the alternative approach proposed by \citet{Barros_etal_2019} which consists in imposing a standard $\Lambda $CDM background expansion history, i.e. setting by construction the {\em constraint}:
\begin{equation}
\label{constraint}
 H^{2}=H^{2}_{\Lambda {\rm CDM}} \,,  
\end{equation}
where $H_{\Lambda {\rm CDM}}$ is the standard Hubble function defined by
\begin{equation}
    H^{2}_{\Lambda {\rm CDM}} = \frac{\kappa ^{2}}{3}(\rho _{r} + \rho _{b} + \rho _{\rm CDM} + \rho _{\Lambda })
\end{equation}
and let this constraint determine an {\em effective} potential $V(\phi )$ according to the resulting dynamics of the scalar field $\phi $. It is important to notice that within such approach $\rho _{\rm CDM}$ and $\rho _{c}$ become two different quantities, as a consequence of the non-standard evolution of the CDM density in the coupled case described by Eq.~(\ref{cdmevolution}).

As the {\em constraint} defined by Eq.~(\ref{constraint}) is the main feature that distinguishes this approach from the standard interacting DE cosmologies, we take the freedom to dub this model as the {\em Constrained Interacting Dark EneRgy} scenario, or {\em CIDER}.
Such constraint (\ref{constraint}) leads to a set of equations characterising the dynamics of the model and its relation to the dynamics of the $\Lambda $CDM cosmology sharing the same expansion history. In particular, by taking the time derivative of Eq.~(\ref{constraint}) and using the continuity Equations (\ref{kleingordon}-\ref{continuity_r}) one gets the scalar field density and pressure as:
\begin{eqnarray}
\label{newenergy}
\rho _{\phi } &=& \rho _{\rm CDM} + \rho _{\Lambda } - \rho _{c}\\
\label{newpressure}
p_{\phi }&=&p_{\Lambda }=-\rho _{\Lambda}
\end{eqnarray}
which can be combined with Eqs.~(\ref{phidensity}-\ref{phipressure}) to get:
\begin{equation}
\label{mainrelation}
    \dot{\phi }^{2} = \rho _{\rm CDM} - \rho _{\phi }\,.
\end{equation}

\begin{table}
    \centering
    \begin{tabular}{cccc}
    \hline
    \hline
    Model & $\beta $ & \makecell{$\sigma _{8}$\\$\mathcal{A}_s=2.105\times 10^{-9}$\\{\em Planck (2020)}} & \makecell{$\sigma _{8}$\\$\mathcal{A}_s=1.992\times 10^{-9}$\\{\em Simulations}}\\
    \hline
    $\Lambda $CDM & 0 & 0.810 & 0.788\\
    CIDER-003 & 0.03 & 0.784 & 0.763\\
    CIDER-005 & 0.05 & 0.743 & 0.723\\
    CIDER-008 & 0.08 & 0.660 & 0.642\\
         \hline
         \hline
    \end{tabular}
    \caption{A summary of the cosmological models investigated in the present work, indicating the value of the only free parameter of th model, i.e. the coupling $\beta$, and the derived value of $\sigma _{8}$ obtained from numerically solving the linear perturbations equations for each model. The table shows both the case of the best-fit {\em Planck 2020} normalisation and the lower normalisation adopted for the simulations presented here.}
    \label{tab:models}
\end{table}

The main outcome of this procedure is that the scalar field potential $V(\phi )$ can be determined a posteriori as a combination of Eqs.~(\ref{newenergy}-\ref{newpressure}) taking the form:
\begin{equation}
\label{potential}
    V(\phi ) = \frac{1}{2}\dot{\phi }^{2} + \rho _{\Lambda }\,,
\end{equation}
and reconstructed along the evolution trajectory of the field, so that the coupling $\beta $ remains the only free parameter of the model.

Interestingly, this represents to some extent a counterpart -- in the realm of Coupled Quintessence cosmologies -- of the widely-investigated \citet{Hu_Sawicki_2007} $f(R)$ gravity model: also in that case, in fact, the free function of the model $f(R)$ is implicitly fixed by the constraint of matching a $\Lambda $CDM expansion history, thereby leaving only one free parameter of the model (the $f_{R0}$ amplitude), and also in that case this constraint makes the model indistinguishable from standard gravity (i.e. $\Lambda $CDM) at the background level. 

Finally, by taking the time derivative of Eq.~(\ref{mainrelation}) one can derive the scalar field equation of motion:
\begin{equation}
\label{field}
    2\ddot{\phi } + 3H\dot{\phi } -\kappa \beta \rho _{c} = 0
\end{equation}
which can be numerically integrated for different values of the coupling $\beta$ to obtain the solution for the dynamical evolution of the system. In particular, we have solved the equation for the models summarised in Table~\ref{tab:models}  assuming the cosmological parameters described in Table~\ref{tab:parameters} and initial conditions consistent with a static field in the very early universe (i.e. deep in the radiation-dominated epoch $\phi _{i}=\dot{\phi}_{i}=0$). 
\begin{table}
    \centering
    \begin{tabular}{cccc}
    \hline
    \hline
    $\Omega _{\rm CDM}$ & $0.262$  & &\\
    $\Omega _{\Lambda} $ & $0.689$ & & \\
    $\Omega _{b}$ & $0.049$ & &\\
    $h$ & $0.677$ & &\\
    $n_{s}$ & $0.9665$ & & \\
    $\mathcal{A}_{s}$ & $2.105\times 10^{-9}$ & $\sigma _{8}=0.810$ & {\em (Planck 2020)}\\
    $\mathcal{A}_{s}$ & $1.992\times 10^{-9}$ & $\sigma _{8}=0.788$ & {\em (Simulations)}\\
         \hline
         \hline
    \end{tabular}
    \caption{The set of cosmological parameters adopted for all the models investigated in the present work.}
    \label{tab:parameters}
\end{table}

\begin{figure*}
\label{fig:background}
\includegraphics[width=\textwidth]{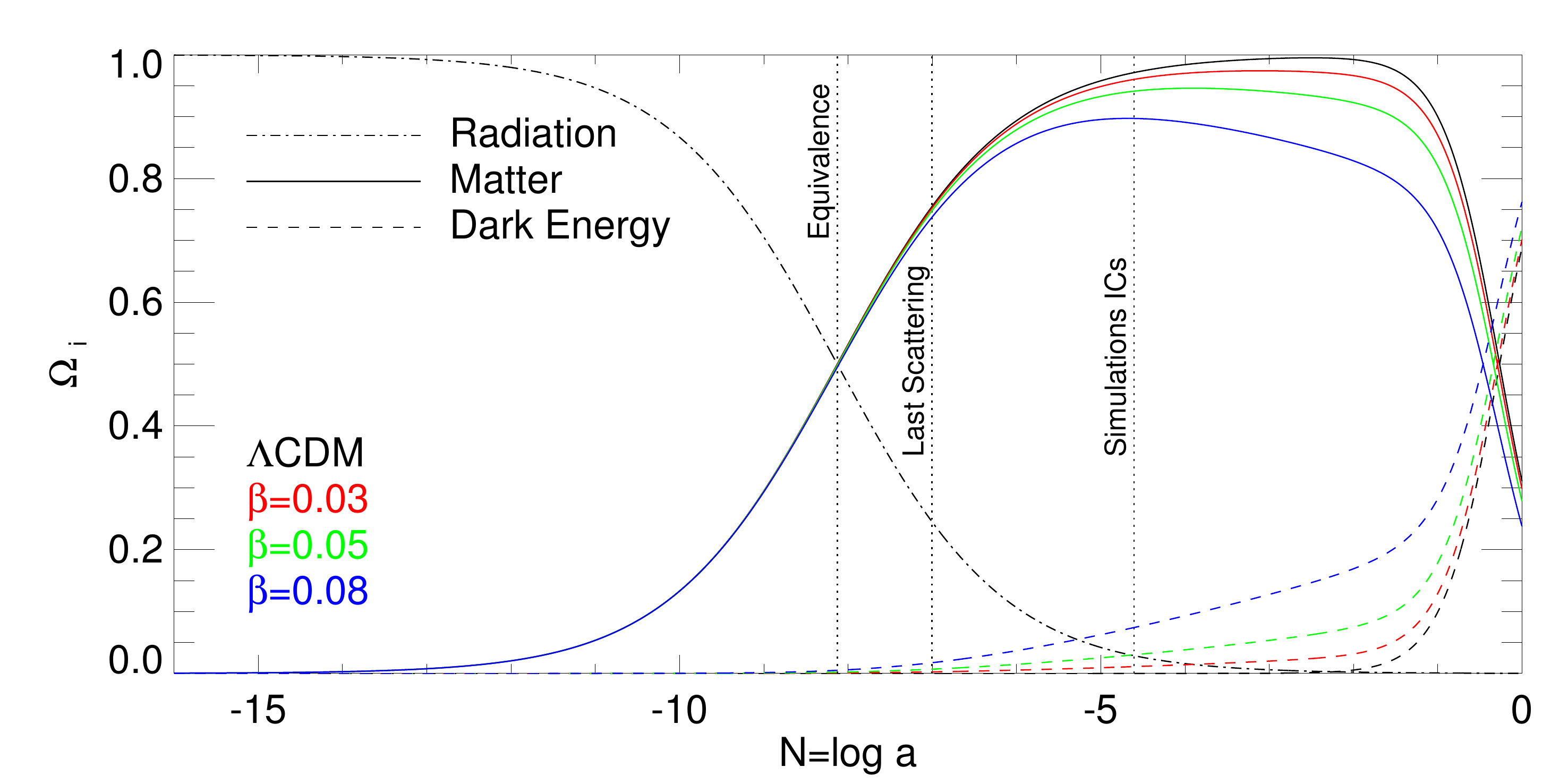}

\caption{The evolution of the radiation (dot-dashed), matter (solid) and DE (dahsed) relative density parameters $\Omega _{i}$ as a function of the e-folding variable $N\equiv \ln a$ for three values of the coupling $\beta = 0.03$ (red) $\beta = 0.05$ (green) and $\beta = 0.08$ (blue) compared with the standard $\Lambda $CDM case (black)}
\label{fig:background}
\end{figure*}

In Fig.~\ref{fig:background} we show the evolution of the different density parameters $\Omega _{i} \equiv \kappa \rho _{i}/(3H^{2})$ for the different components ({\em dot-dashed} for radiation, {\em solid} for total matter i.e. CDM+baryons, and {\em dashed} for DE) as a function of the number of e-folds $N\equiv \ln(a)$ for these three values of the coupling and for the standard $\Lambda $CDM cosmology. As one can clearly see from the plot, increasing the value of the coupling leads to a suppression of the $\Omega _{c+b}$ density in the matter-dominated epoch, with a corresponding increase of the $\Omega_{\phi }$ density. However, it should be stressed again that such deviations cannot be observed in any way through geometric probes as the background expansion history remains identical to the standard $\Lambda $CDM case by construction. Therefore, the standard bounds based e.g. on Early Dark Energy constraints \citep[e.g.][]{Riess_etal_2006} or on other background observables \citep[as e.g. in][]{Caprini_Tamanini_2016} do not apply here. For the common coupling values, our results are consistent with the findings of \citet{Barros_etal_2019} (see e.g. their Fig.~1).

\begin{figure*}
\label{fig:phi}
\includegraphics[width=\columnwidth]{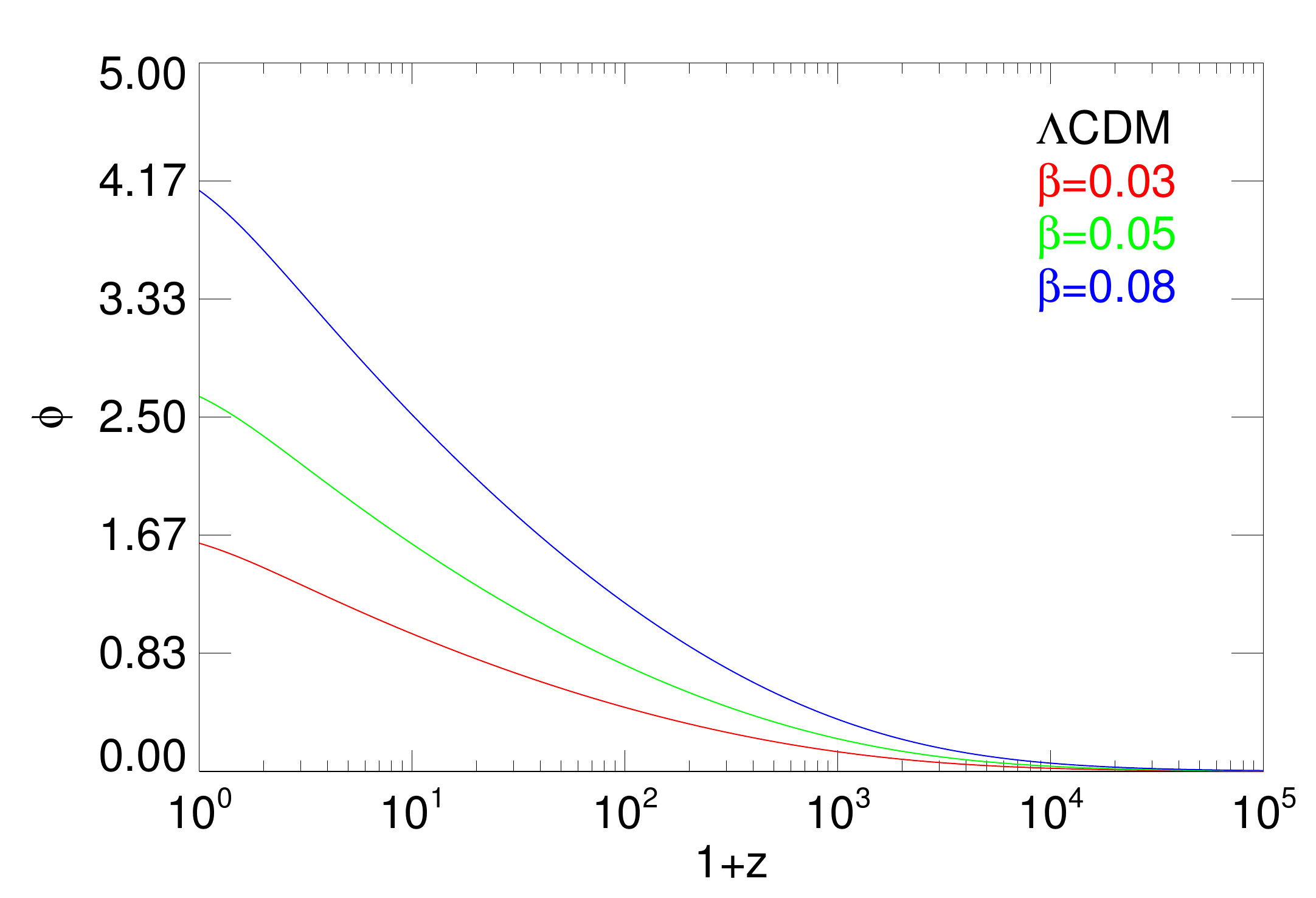}
\includegraphics[width=\columnwidth]{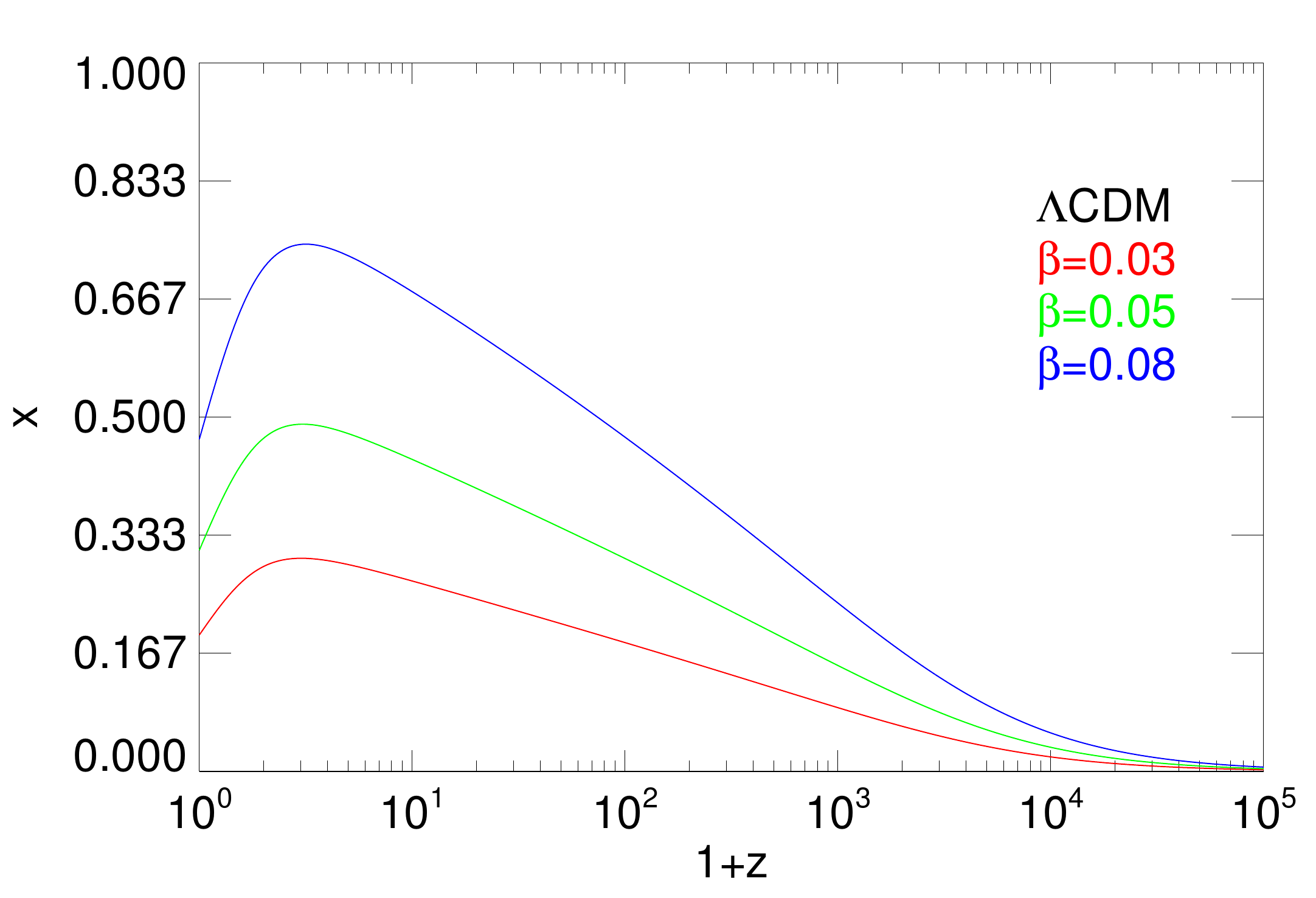}\\
\includegraphics[width=\columnwidth]{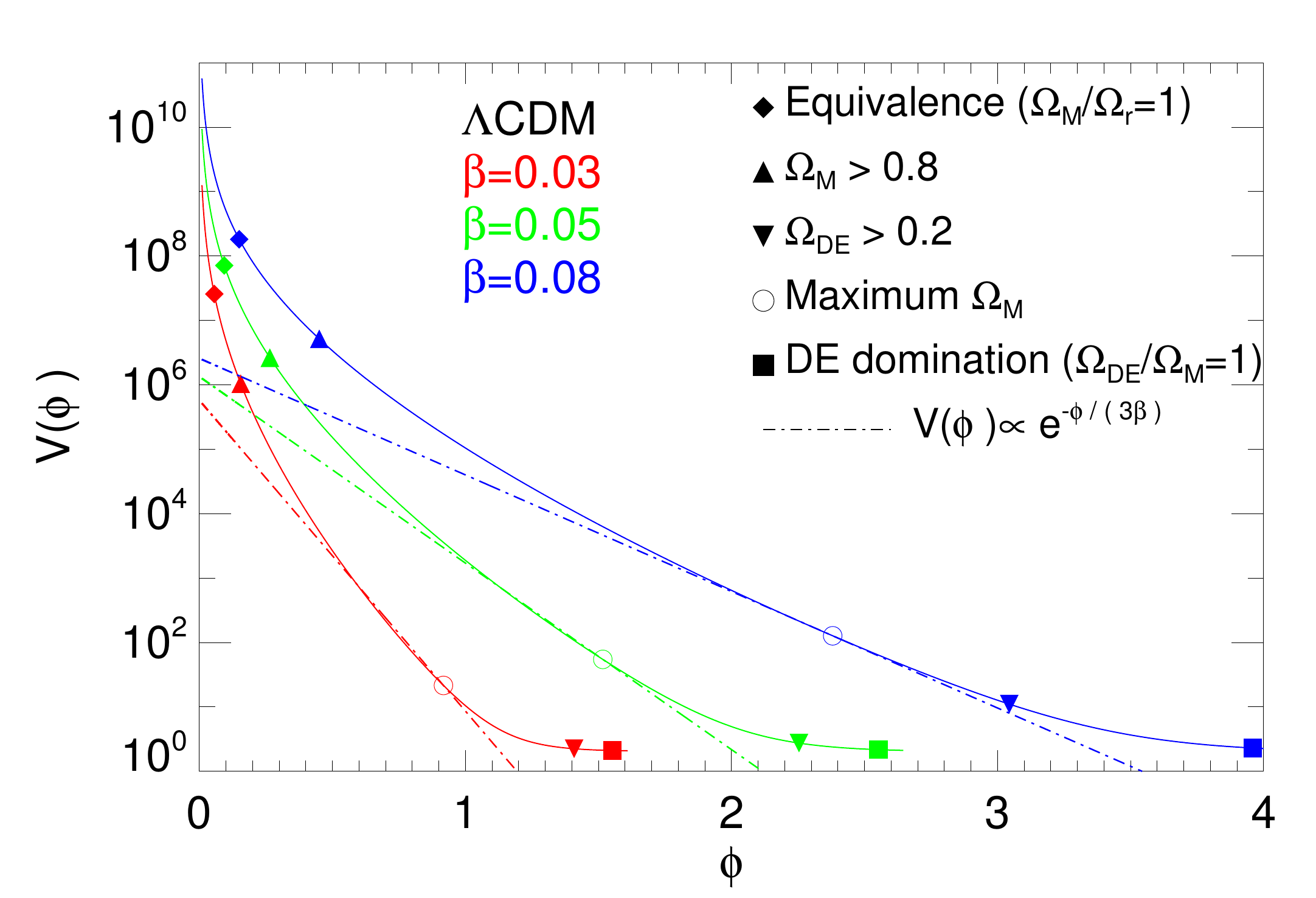}
\includegraphics[width=\columnwidth]{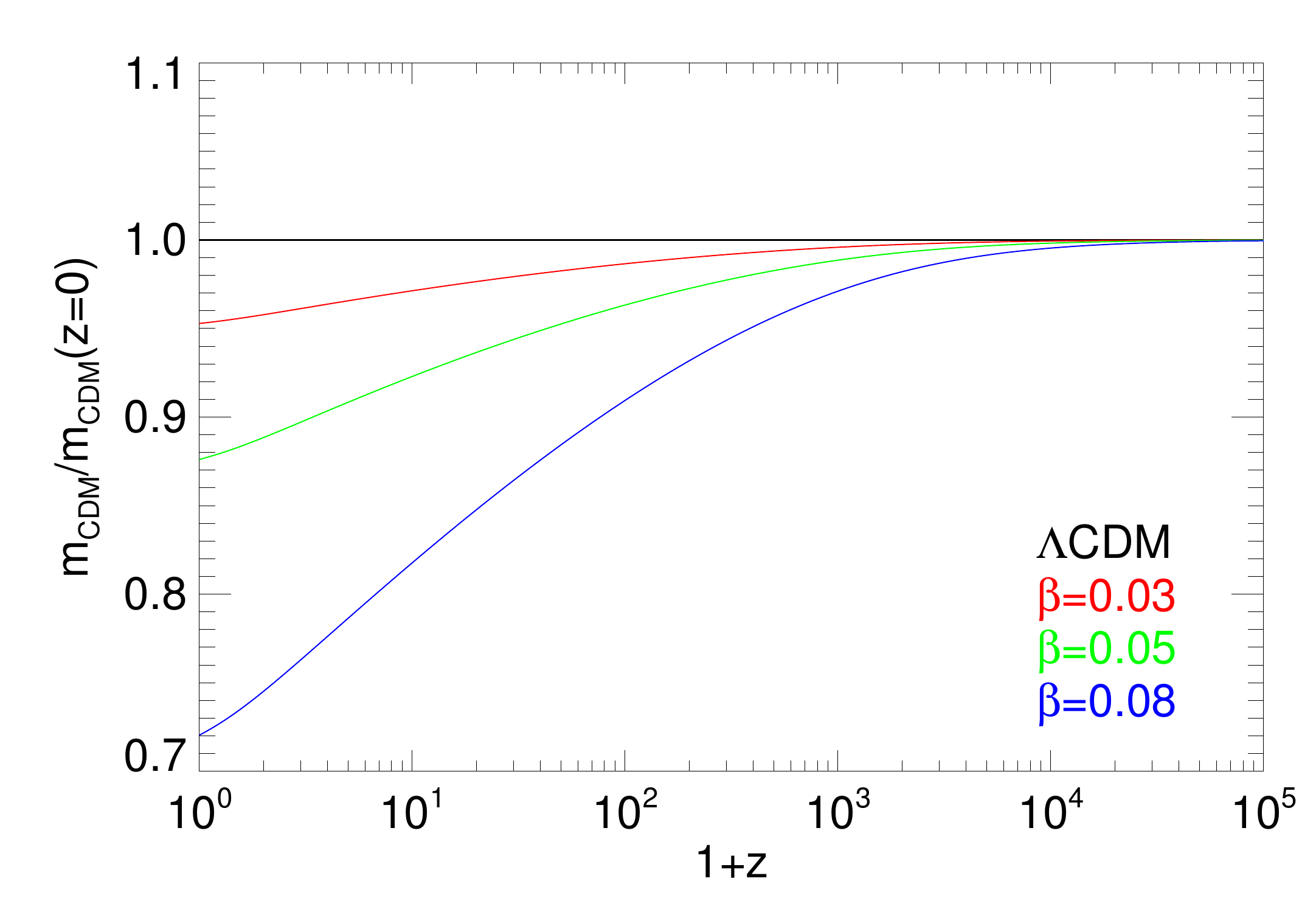}

\caption{The evolution of some relevant background quantities: {\em (top-left):} the scalar field evolution as a function of redshift; {\em (top-right):} the scalar field kinetic term $x\equiv \sqrt{\kappa /6}(\dot{\phi }/H)$ as a function of redshift; {\em (bottom-right):} the CDM particle mass as a function of redshift; {\em (bottom-left):} the reconstructed scalar field potential obtained from the numerical solution of Eq.~\ref{potential} showing the asymptotic slope of $1/(3\beta )$ during matter domination.}
\label{fig:bkgdproperties}
\end{figure*}

In Fig.~\ref{fig:bkgdproperties} we also show the evolution of some characteristic quantities as obtained from our numerical integration of Eq.~\ref{field}. More specifically, in the top-left panel we display the evolution of the scalar field value $\phi $ (in units of the reduced Planck mass) as a function of redshift, for the three couplings under investigation. As the figure shows, the field remains static at very high redshifts and starts moving towards positive values under the effect of the source term provided by the non-vanishing coupling in Eq.~\ref{field}. The field displacement at $z=0$ is roughly proportional to the value of the coupling.
In the top-right panel we show the kinetic term of the scalar field defined as $x\equiv \sqrt{\kappa /6}(\dot{\phi }/H)$. Also in this case, the field shows -- at any redshift -- a larger kinetic energy for larger values of the coupling, and the decrease of $x$ at low redshifts is related to the transition between matter and DE domination.

In the bottom-left panel, we show the interesting outcome of the {\em a-posteriori} reconstruction of the field self-interaction potential $V(\phi )$, according to Eq.~\ref{potential} and to the numerical solution of the field dynamics. As one can see, the reconstructed potential resembles an exponential with an evolving slope as a function of the field $\phi $. In particular, we show that at the time of maximum matter relative density -- that is to say when $\Omega _{\rm M}$ reaches its maximum value over the whole cosmic expansion history (marked by an open circle on the different trajectories in the figure) -- the potential is well approximated by an exponential of the form 
\begin{equation}
  V_{\rm eff}(\phi ) \propto \exp[-\phi /(3\beta )]\,  
\end{equation}
which remains a decent approximation during most of the matter dominated epoch. The figure also shows some relevant cosmic times along the evolution of the scalar field on its own self-interacing potential.

Finally, in the bottom-right panel of Fig.~\ref{fig:bkgdproperties} we show the evolution of the CDM particle mass due to the energy-momentum exchange with the DE scalar field. As one can see from the plot, the CDM particle mass decreases in time (consistently with the dimensionless density evolution depicted in Fig.~\ref{fig:background}), reaching a suppression of about $30\%$ for the strongest coupling under investigation. This behaviour will be responsible, as we will see below, for the suppressed growth of density perturbations at low redshifts compared to the standard $\Lambda $CDM reference model. 

\subsection{Linear perturbations}
\label{sec:linear}

\begin{figure*}
    \centering
    \includegraphics[width=0.49\textwidth]{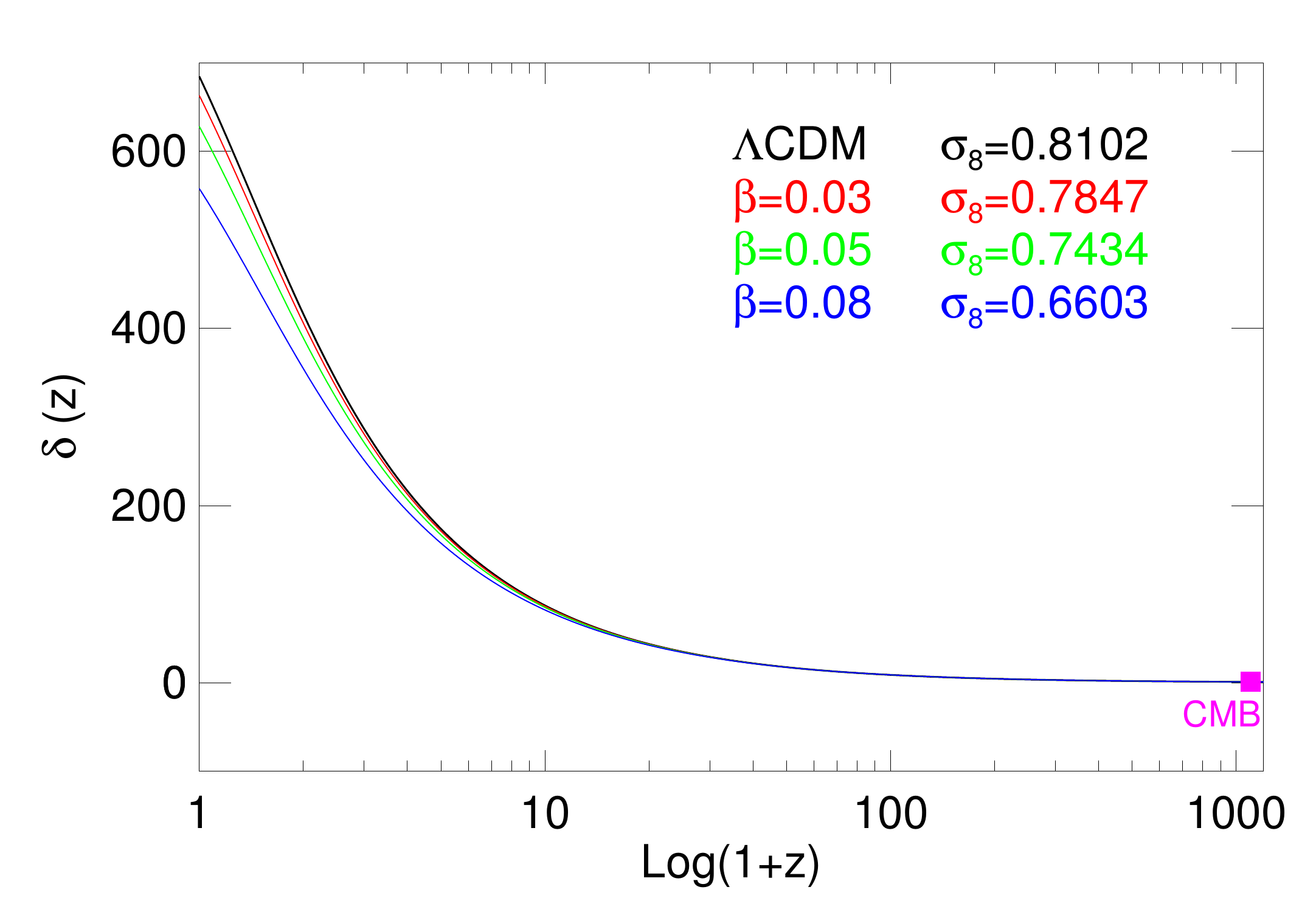}
    \includegraphics[width=0.49\textwidth]{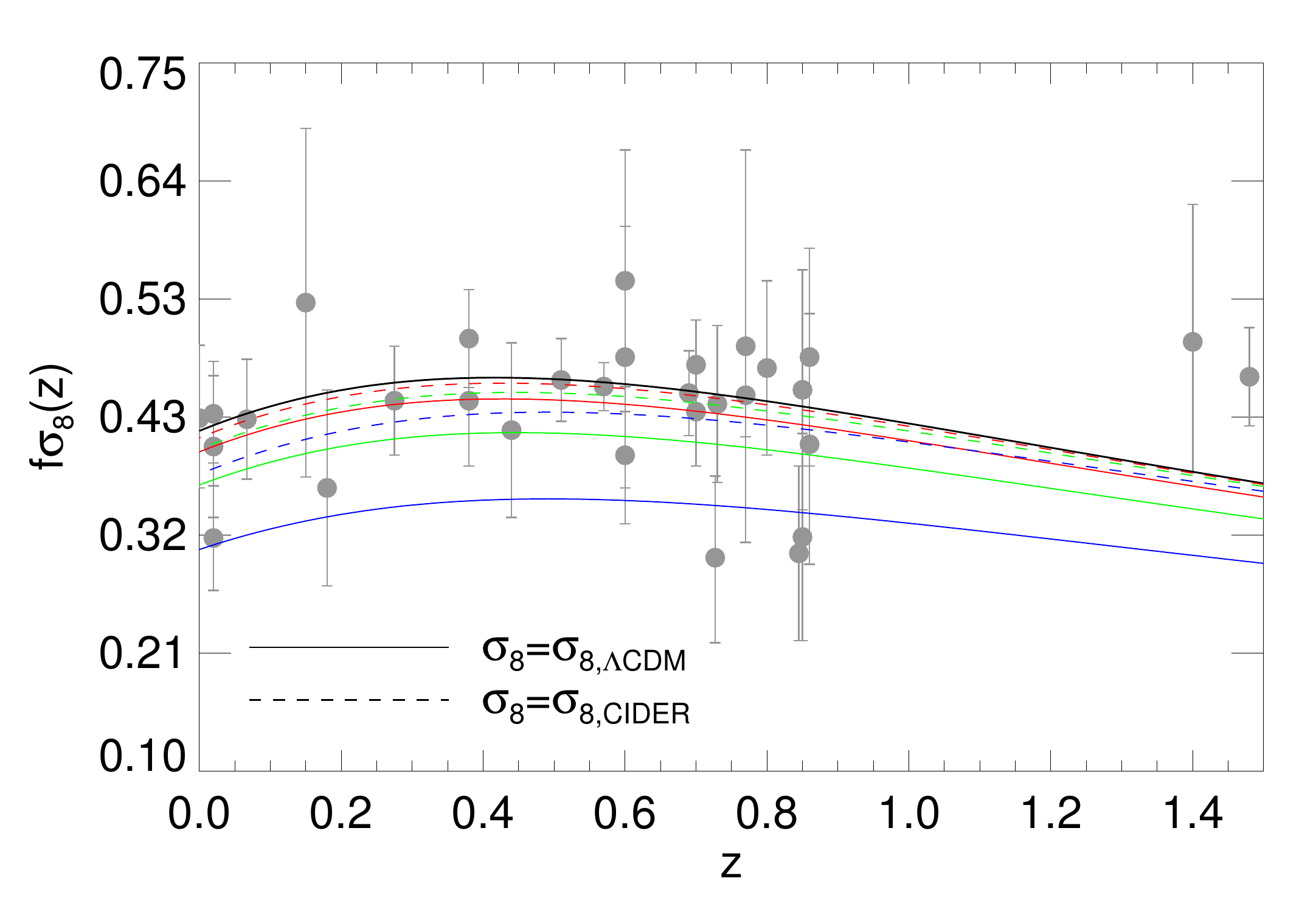}
    \caption{{\em Left panel:} The redshift evolution of the linear density growing mode as computed through a numerical integration of Eqs.~(\ref{pertc}-\ref{pertb}) for the different models under investigation. The legend indicates the derived value of $\sigma _{8}$ for each model assuming the same normalisation of scalar perturbations $\mathcal{A}_{s}$ at the last scattering surface; {\em Right panel:} The comparison of the derived quantity $f\sigma _{8}$ as a function of redshift with a compilation of observational data collected in \citet{Marulli_etal_2021} assuming the $\Lambda $CDM fiducial value for $\sigma _{8}$ for all the models \citep[as done in ][ solid curves]{Barros_etal_2019} or by adopting the true value of $\sigma _{8}$ as indicated in Table~\ref{tab:models} and in the {\em left} panel (dashed  curves).}
    \label{fig:linear_integration}
\end{figure*}

We now move to discuss the evolution of linear perturbations in the CIDER scenario, on top of the background evolution described in the previous section. We consider a perturbed FLRW metric in the Newtonian gauge:
\begin{equation}
    ds^{2} = -(1+2\Psi )dt^{2} - a^{2}(t)(1-2\Phi )\delta _{ij}dx^{i}dx^{i}
\end{equation}
and perturbed fields and fluids in the form
\begin{eqnarray}
\phi (\vec{x},t) &=& \bar{\phi }(t) + \delta \phi (\vec{x},t)\\
\rho _{i}(\vec{x},t) &=& \bar{\rho }_{i} + \delta \rho _{i}(\vec{x},t)\\
p_{i}(\vec{x},t) &=& \bar{p}_{i} + \delta p_{i}(\vec{x},t)
\end{eqnarray}
for which we can define in the usual way the density contrast $\delta _{i} \equiv \delta \rho _{i}/\bar{\rho} _{i}$ of each species.

By plugging these definitions into the perturbed Einstein equations one can derive the dynamic equations for the density contrast of the different species, and in particular for the DM and the baryonic components, that with some algebra and by considering the Newtonian limit (i.e. sub-horizon scales of the perturbations) can be written as \citep[see][for a more detailed derivation]{Amendola_2004,Baldi_2011a}:
\begin{eqnarray}
&\delta _{c}''+\delta _{c}'\left( 2+\frac{H'}{H}-\kappa \beta \phi '\right) \nonumber \\
&-\frac{3}{2}\delta _{c}\left( \Omega _{\rm CDM} -\frac{\kappa ^{2}}{3}\phi '^{2}\right)(1+2\beta ^{2}) - 
\label{pertc}
\frac{3}{2}\Omega _{b}\delta _{b} = 0 \\
&\delta _{b}''+\delta _{b}'\left( 2+\frac{H'}{H}
\right) \nonumber \\
\label{pertb}
&-\frac{3}{2}\delta _{c}\left( \Omega _{\rm CDM} -\frac{\kappa ^{2}}{3}\phi '^{2}\right) - \frac{3}{2}\Omega _{b}\delta _{b} = 0 \,,
\end{eqnarray}
where a prime indicates a derivative with respect to the e-folding variable $N$.
It is interesting to notice how both baryons and DM components are subject to a DM source term with an effective density given by the $\Lambda $CDM value $\Omega _{\rm CDM}$ modulated by the particle mass suppression $\kappa ^{2} \beta \phi '^{2}/3$, while only the DM component is subject to the additional fifth-force term $\propto(1+2\beta ^2)\delta _{c}$ and friction term $\propto -\kappa \beta \phi ' \delta _{c}'$. This is a well-known feature of coupled DE models \citep[see e.g.][]{Mainini_Bonometto_2004,Baldi_etal_2010}.

Eqs.~(\ref{pertc}-\ref{pertb}) can be numerically integrated to obtain the linear growth of density perturbations for the DM and baryonic components, respectively, which can then be combined to obtain the growth of the total matter density contrast defined as:
\begin{equation}
    \delta \equiv \frac{\Omega _{c}\delta _{c} + \Omega _{b}\delta _{b}}{\Omega _{c}+\Omega _{b}}\,.
\end{equation}

In Fig.~\ref{fig:linear_integration} we show the results of the numerical integration of the linear perturbations equations (\ref{pertc}-\ref{pertb}), displaying in particular in the left panel  the evolution as a function of redshift of the total density contrast $\delta $ normalised at the redshift of the Cosmic Microwave Background ($z_{\rm CMB}\approx 1100$) and in the right panel the linear growth observable $f\sigma _{8}$ -- where $f\equiv d \ln \delta /dN$ is the linear growth rate -- compared with observational data obtained from different surveys as described in  \citet[][]{Marulli_etal_2021} (and references therein). These two plots correspond to the ones displayed in \citet[][]{Barros_etal_2019} in their Figs.~3 and 2, respectively, and appear fully consistent with their results. In the latter plot, we also distinguish between the case where $\sigma _{8}$ is assumed to take the $\Lambda $CDM fiducial value \citep[as done in ][ solid curves]{Barros_etal_2019} and the case where $f\sigma _{8}$ is computed for every model by using its actual $\sigma _{8}$ value (dashed curves) as listed in Table~\ref{tab:models} and indicated in the figure legend. As one can see in the plots, a larger value of the coupling $\beta $ determines a stronger suppression of the linear growth and a lower $f\sigma _{8}$, thereby resulting in a lower value of $\sigma _{8}$ at $z=0$. This is the reason why these models have been claimed to provide a possible solution to the persisting $\sigma _{8}$ tension (see discusson and references above), and certainly represents a particularly appealing feature of this scenario as a suppression in the growth of structures is difficult to achieve in most DE or Modified Gravity scenarios, though with some noticeable exceptions \citep[see e.g.][]{Pourtsidou_Tram_2016,Baldi_Simpson_2017,Wittner_etal_2020}.

\begin{figure*}
\includegraphics[width=0.49\textwidth]{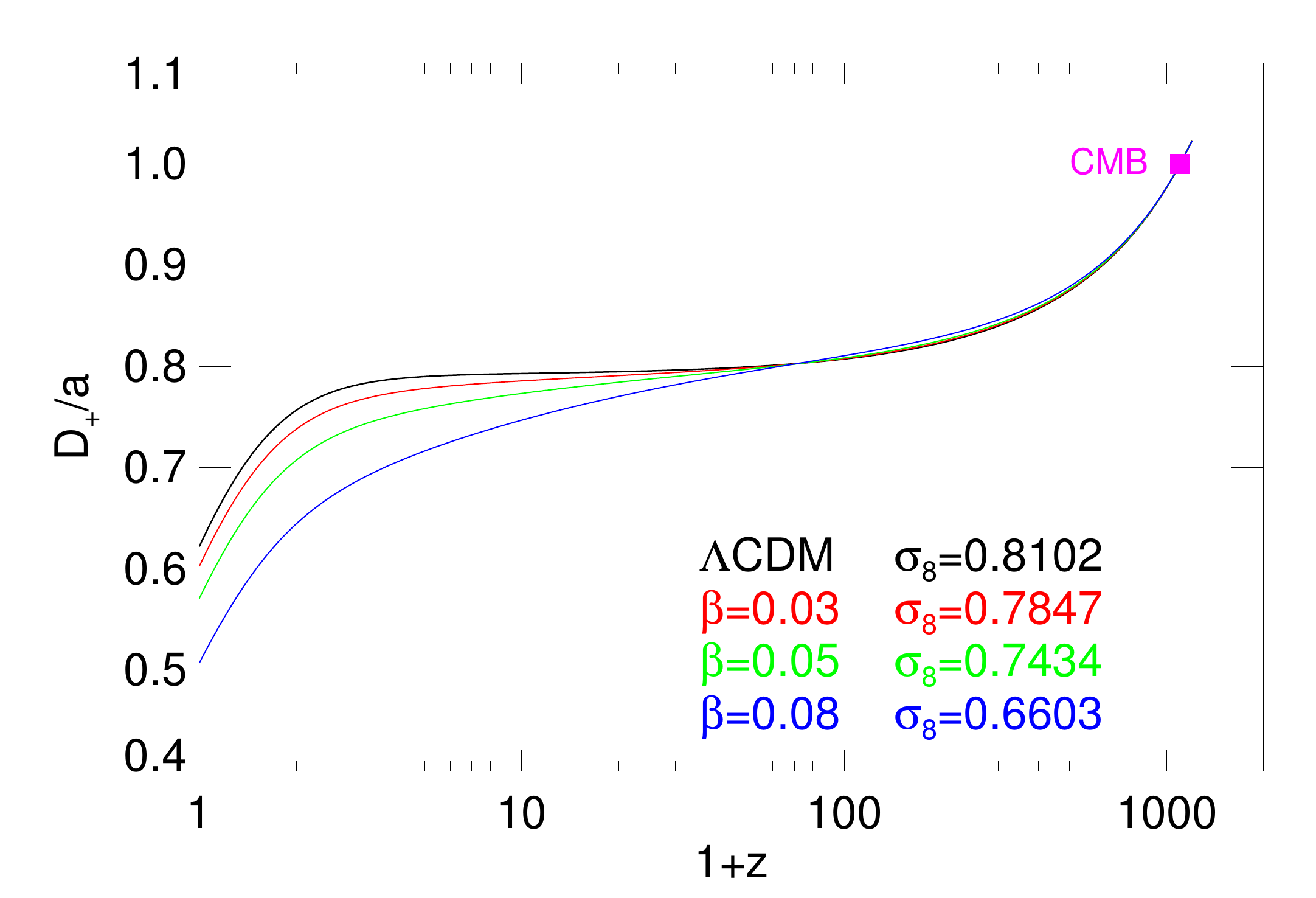}
\includegraphics[width=0.49\textwidth]{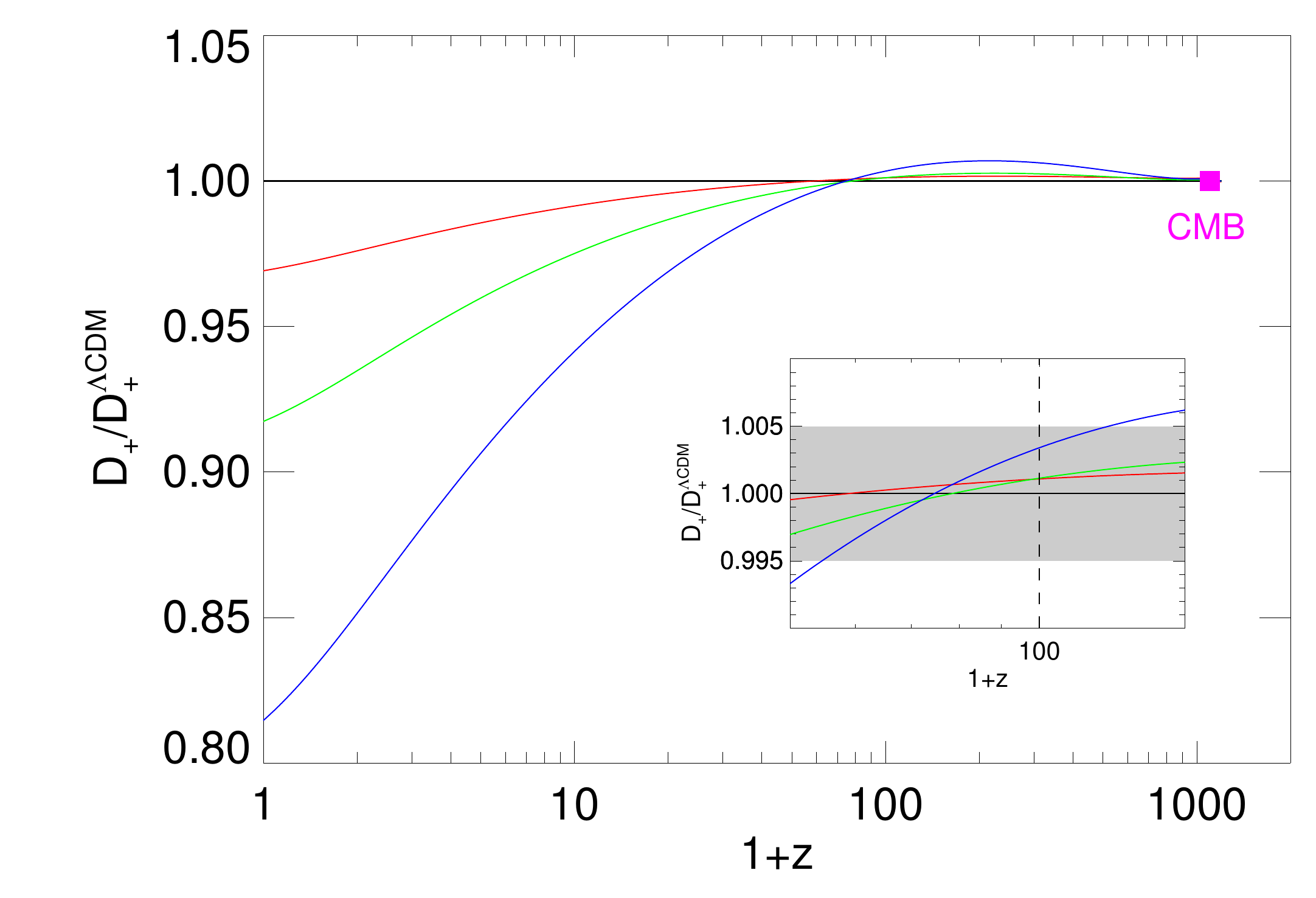}
\caption{{\em Left}: The linear growth factor $D_{+}$ in units of the cosmological scale factor $a$ normalised at the redshift of the CMB $z_{\rm CMB}\approx 1100$ for all the models under investigation. The slower growth of linear perturbations in the CIDER cosmologies results in a lower value of $\sigma _{8}$ at low redshifts. {\em Right}: The ratio of the linear growth factor $D_{+}$ to the reference $\Lambda $CDM case for all the CIDER models. The inset shows the evolution of the ratio around the starting redshift of our simulations at $z_{i}=99$. All models have amplitudes within $0.5\%$ of the reference model at $z_{i}$.}
\label{fig:linear}
\end{figure*}

In Fig.~\ref{fig:linear} we finally display the linear growth factor and its ratio to the $\Lambda $CDM reference model, both normalised at $z_{\rm CMB}$, in the left and right panels, respectively. In particular, the latter figure allows to notice how at high redshifts (roughly in the range $70 < z < z_{\rm CMB}$) the coupling induces a slightly enhanced growth of the total density contrast, followed by a more significant suppression of the growth at lower redshifts. The early enhanced growth is driven by the cumulative effects of the extra-friction and fifth-force terms in Eq.~(\ref{pertc}) before the DM mass decrease starts to dominate and to suppress the growth at later stages. In the inset of the right plot this can be seen in more detail, also noticing that all the models cross the reference $\Lambda $CDM line at $z\approx 70$, and that in the redshift range $60 < z < 100$ all models have perturbations amplitudes within $0.5\%$ from the reference $\Lambda $CDM. This finding will be relevant to support our approach to define the initial conditions of the simulations that will be discussed in the next Section.

\section{The simulations}
\label{sec:sims}

\begin{figure*}
\includegraphics[width=0.49\textwidth]{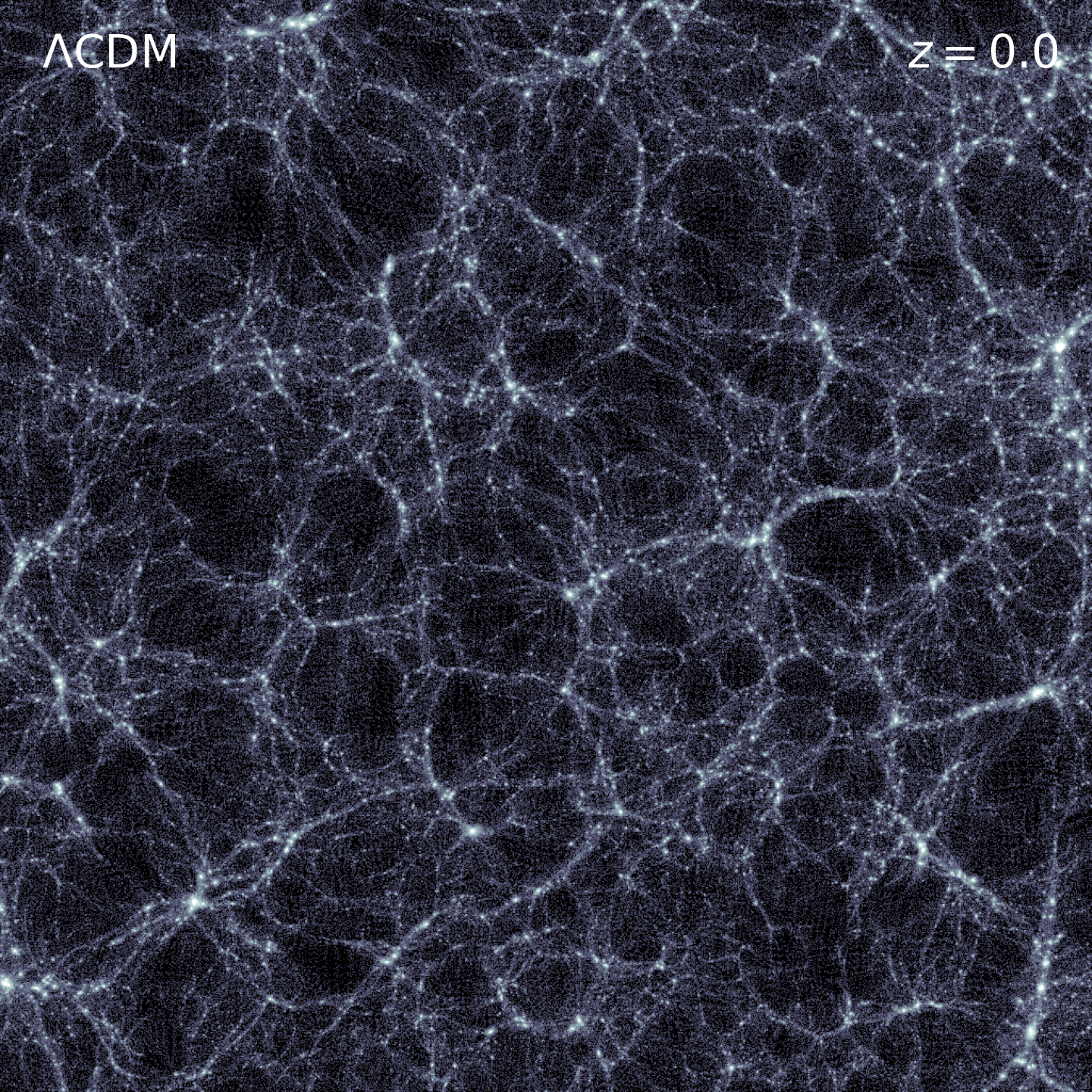}
\includegraphics[width=0.49\textwidth]{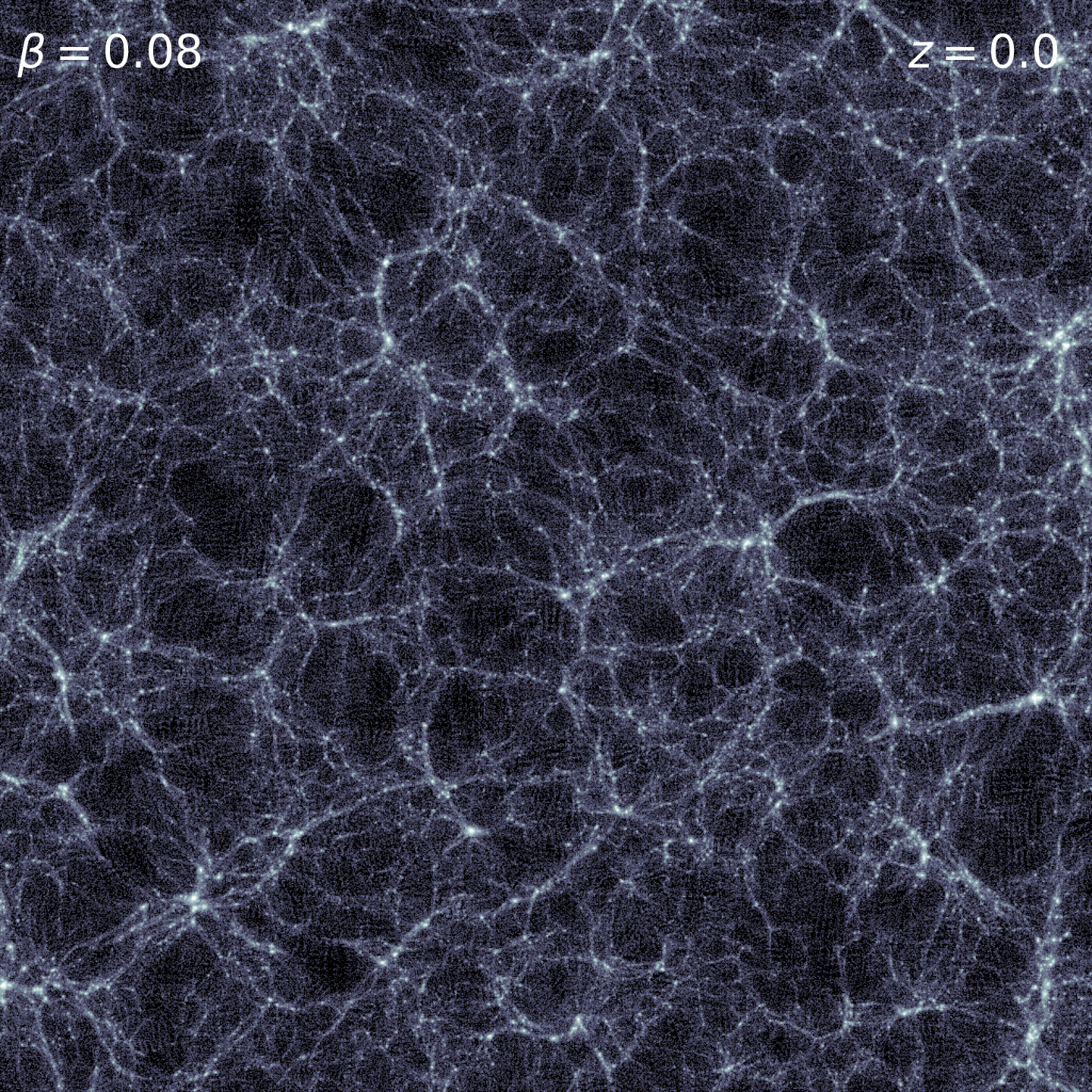}
\label{fig:mass_maps}
\caption{The density distribution in a slice of side $300$ Mpc$/h$ and thickness $20$ Mpc$/h$ extracted from the cosmological box of the $\Lambda $CDM ({\em left}) and $\beta = 0.08$ CIDER ({\em right}) cosmologies. The differences in the distribution and height of the density peaks can be noticed by eye by looking e.g. at the structures appearing in the bottom-right corner of the two figures.}
\end{figure*}

\subsection{The CoDECS2 Project}

The {\small CoDECS} project \citep[][]{CoDECS} was developed about 10 years ago as a publicly available suite of cosmological simulations for different types of interacting Dark Energy models, with the aim to provide a large dataset of simulated cosmologies spanning the (at the time) viable range of interacting DE scenarios, including different self-interaction potentials and coupling functions. Most of the models included in the original {\small CoDECS} suite have subsequently been tightly constrained by data \citep[see again][]{Pettorino_etal_2012,Pettorino_2013,Planck_2015_XIV,Gomez-Valent_Pettorino_Amendola_2020} and appear today less appealing. However, new interacting DE flavours have emerged in the meantime -- including e.g. the CIDER scenario discussed in the present work \citep[][]{Barros_etal_2019}, or other forms of time-dependent couplings \citep[e.g.][]{VanDeBruck_Thomas_2019}, or models of pure momentum exchange \citep[][]{Simpson_2010,Pourtsidou_etal_2013,Skordis_Pourtsidou_Copeland_2015}, or multiple coupling scenarios \citep[as e.g.][]{Bonometto_etal_2015,Baldi_2012a,Vagnozzi_etal_2020,Ferlito_etal_2022} -- that appear to be less constrained by data and may provide an interesting phenomenology, possibly alleviating some of the currently unaccounted tensions between high-redshift and low-redshift observations \citep[][]{Pourtsidou_Tram_2016,Baldi_Simpson_2017} or persisting problems of the standard $\Lambda $CDM model at small scales \citep[][]{Maccio_etal_2015,Garaldi_Baldi_Moscardini_2016}. 
This motivates a follow-up of the {\small CoDECS} project by running a new suite of simulations for this new generation of interacting DE cosmologies. We are currently developing such a follow-up as the {\small CoDECS2} project, of which the {\small CIDER} simulations discussed in this work represent the first step.

\subsection{The CIDER simulations suite}

We performed a suite of collisionless cosmological N-body simulations for the set of CIDER models liested in Table~\ref{tab:models} using the {\small C-Gadget} code \citep[][]{Baldi_etal_2010}, a modified version of the {\small GADGET3} code \citep[][]{gadget-2} that implements the type of dark sector interactions characterising general Coupled DE models, which can be straightforwardly applied to the CIDER scenario. {\small C-Gadget} allows to follow the cosmological evolution of different particle species featuring different couplings with a light DE scalar field, thereby allowing to distinguish the effective interactions experienced by uncoupled baryons and coupled DM particles and to capture more accurately the real evolution of such models, including the possible segregation effects between these different components. 

Following the approach of the original {\small CoDECS} simulations \citep[][]{CoDECS}  we start the investigation of the non-linear evolution of the CIDER scenario by simulating a cosmological comoving volume of $1 $ Gpc$^{3}h^{-3}$ filled with a collection of $2\times 1024^{3}$ particles to sample the baryons and DM density field. 

The initial conditions have been generated at $z_{s}=99$ as a random-phase realisation of the total matter power spectrum computed by the Boltzmann code {\small CAMB} \citep[][]{camb} for a $\Lambda $CDM cosmology with parameters defined as in Table~\ref{tab:parameters}, according to 2-nd order Lagrangian perturbation theory, by means of the public initial conditions generation code {\small MUSIC} \citep[][]{Music}. 
For these simulations, however, we adopt a lower value of the scalar  perturbations amplitude $\mathcal{A}_{s}$ -- and consequently a lower value of $\sigma _{8}=0.788$ -- with respect to the current $\Lambda $CDM best-fit (as obtained e.g. by \citet{Planck_2018_VI}$, \sigma_{8}=0.811$). This choice is made in order to ensure a reasonable value of $\sigma_8$ for most of the models considered within the range of the {\small CoDECS2} project, as the same $\Lambda $CDM run is assumed as reference also for different types of interacting Dark Energy models included in {\small CoDECS2} and  characterised by an enhanced growth of perturbations. These simulations will be presented in two separate companion papers (Baldi et al. in prep). In any case, the choice of a slightly lower perturbations amplitude for the $\Lambda $CDM reference model does not affect the main conclusions of the present work, which are expected to hold unaltered for a different $\sigma _{8}$ normalisation as well.

Although in general the modified growth of perturbations in coupled DE cosmologies would require to scale the amplitude of particle displacements and velocities in the initial conditions according to the individual growth history of each model (as computed by the numerical integration of Eqs.~\ref{pertc}-\ref{pertb} and depcited in Figs.~\ref{fig:linear_integration}-\ref{fig:linear}), we choose to employ the same $\Lambda $CDM initial conditions for all models based on the observation that -- at the starting redshift of the simulations -- all models have a total linear perturbations amplitude within $0.5\%$ from the reference $\Lambda $CDM cosmology (see the inset in Fig.~\ref{fig:linear} above), thereby making any effect arising from such difference reasonably sub-dominant compared to the subsequent evolution of the simulations.
The use of the {\small MUSIC} code for the generation of initial conditions will allow to perform detailed zoom-in simulations of individual halos extracted from this suite to test the effect of the coupling on the virialisation process of collapsed objects. This will be investigated in a future work.

For all our simulations we stored 25 full particle snapshots and 93 halo catalogues (see below for the details of the halo finding procedure) in the redshift range $0<z<99$, of which 12 and 52, respectively, are covering the redshift range $0<z<1$.
In Figure~\ref{fig:mass_maps} we display the mass density distribution in a slice of thickness $20$ Mpc$/h$ and $300$ Mpc$/h$ per side at $z=0$ for the standard $\Lambda $CDM cosmology ({\em left}) and for the most extreme CIDER model ($\beta = 0.08$, {\em right}). Visual inspection of these two maps allows to clearly identify differences in the spatial distribution and prominence of the density peaks corresponding to collapsed halos, e.g. by looking at the structures appearing in the bottom-right corner of the two images, showing a more evolved and more strongly peaked distribution for the $\Lambda $CDM cosmology. This trend will be quantified through the analysis of the matter density field statistics in Section~\ref{sec:results}.

\subsection{The halo catalogues}

In order to obtain our 93 halo catalogues we have identified collapsed structures by means of a halo finding process based on a Friends-of-Friends algorithm \citep[FoF, ][]{Davis_etal_1985} with linking length $\lambda = 0.2\times \bar{d}$, with $\bar{d}$ being the mean inter-particle separation, followed by a gravitational un-binding procedure of all FoF groups performed with the {\small SubFind} algorithm \citep[][]{Springel_etal_2001}, retaining only bound substructures with a minimum of 20 particles, and computing -- for the substructure hosting the most bound particle -- the spherical overdensity mass $M_{200,x}$ and radius $R_{200,x}$ defined in terms of an overdensity threshold of 200 times the critical density ($\rho _{crit} \equiv 3H^{2}/\kappa ^{2}$) or the mean density ($\rho _{mean}=\Omega _{\rm M}\times \rho _{crit}$) of the universe :
\begin{equation}
    \frac{4}{3}\pi R_{200,\left\{ crit,mean\right\}}^{3}\times 200 \times \rho _{\left\{ crit,mean\right\} } = M_{200,\left\{crit,mean\right\}}\,.
\end{equation}
We are therefore equipped with two different halo catalogues based on these two different choices of the overdensity threshold, that will be employed in our analysis below.

\subsection{Voids identification and selection}
\label{sec:voidfinding}

We also identify cosmic voids in the DM distribution of our simulations snapshots at $z=\left\{ 0, 1\right\}$ by means of the {\small VIDE} toolkit \citep[][]{VIDE}, a void finding algorithm based on a Voronoi tessellation of the simulation domain and on the Watershed Transform \citep[][]{Platen_etal_2007} approach previously implemented in the {\small ZOBOV} code \citep[][]{Neyrinck_2008}. We refer to the original papers of {\small VIDE} and {\small ZOBOV} for a more extended description of the respective algorithms. In the void identification process we allow for a random subsampling of the DM particle ensemble down to an average particle density of $0.02\, h^{3}/$Mpc$^{3}$ to save computational resources, and we clean the resulting catalogue from unphysical voids following the procedure detailed in Section 3 of \citet{Pollina_etal_2016}. For the resulting catalogue, we associate to each void an effective radius $R_{\rm eff}$ defined from the total void volume $V_{\rm VOID}$ detected by {\small VIDE} through the Watershed Transform procedure as
\begin{equation}
\label{effective_radius}
    V_{\rm VOID} = \frac{4}{3}\pi R_{\rm eff}^{3}\,.
\end{equation}

\section{Results}
\label{sec:results}

In the present Section we describe the main results obtained from our suite of simulations with respect to several different cosmological observables.

\subsection{Large-scale matter distribution}

\subsubsection{The matter power spectrum}
\begin{figure*}
\includegraphics[width=0.49\textwidth]{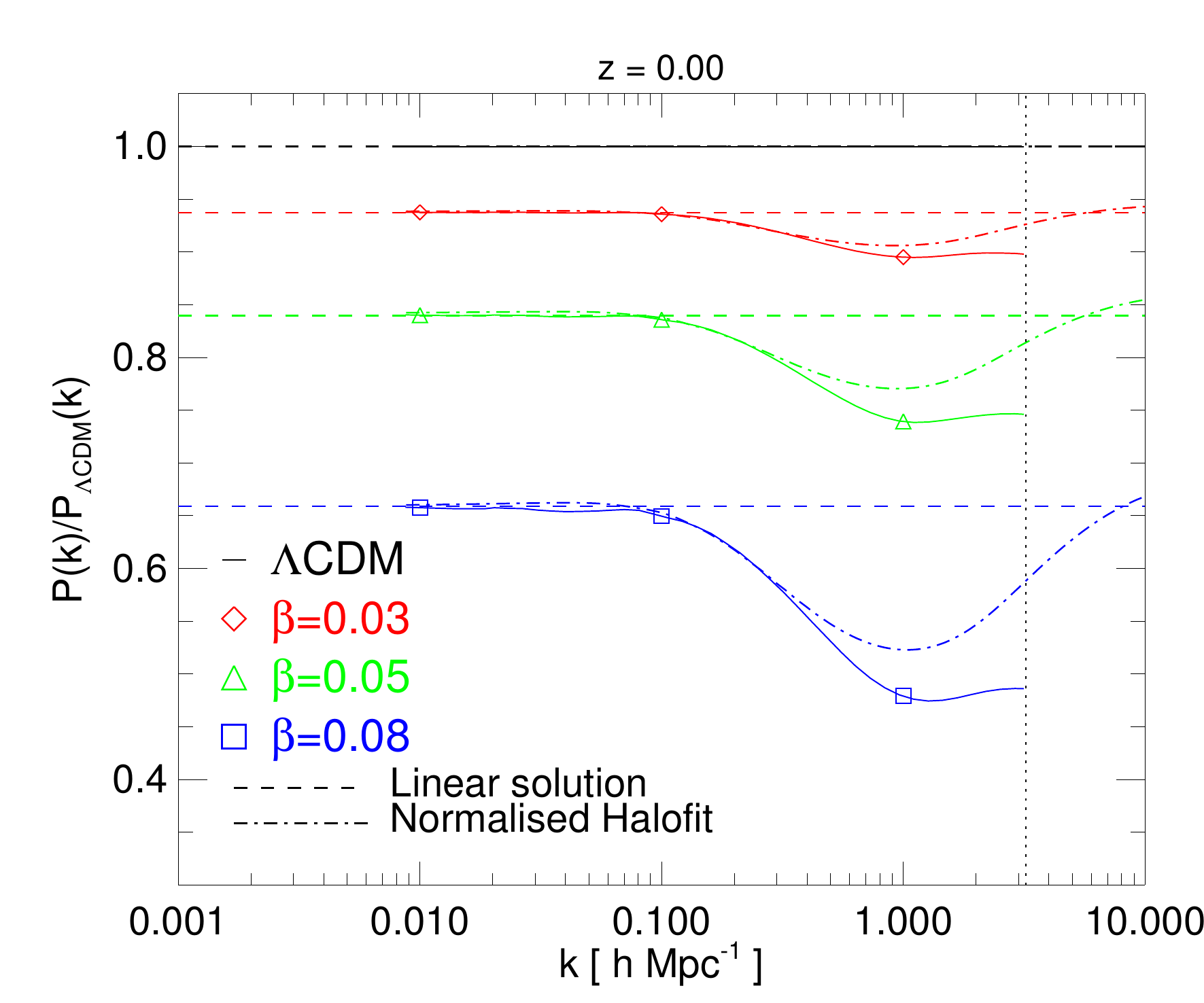}
\includegraphics[width=0.49\textwidth]{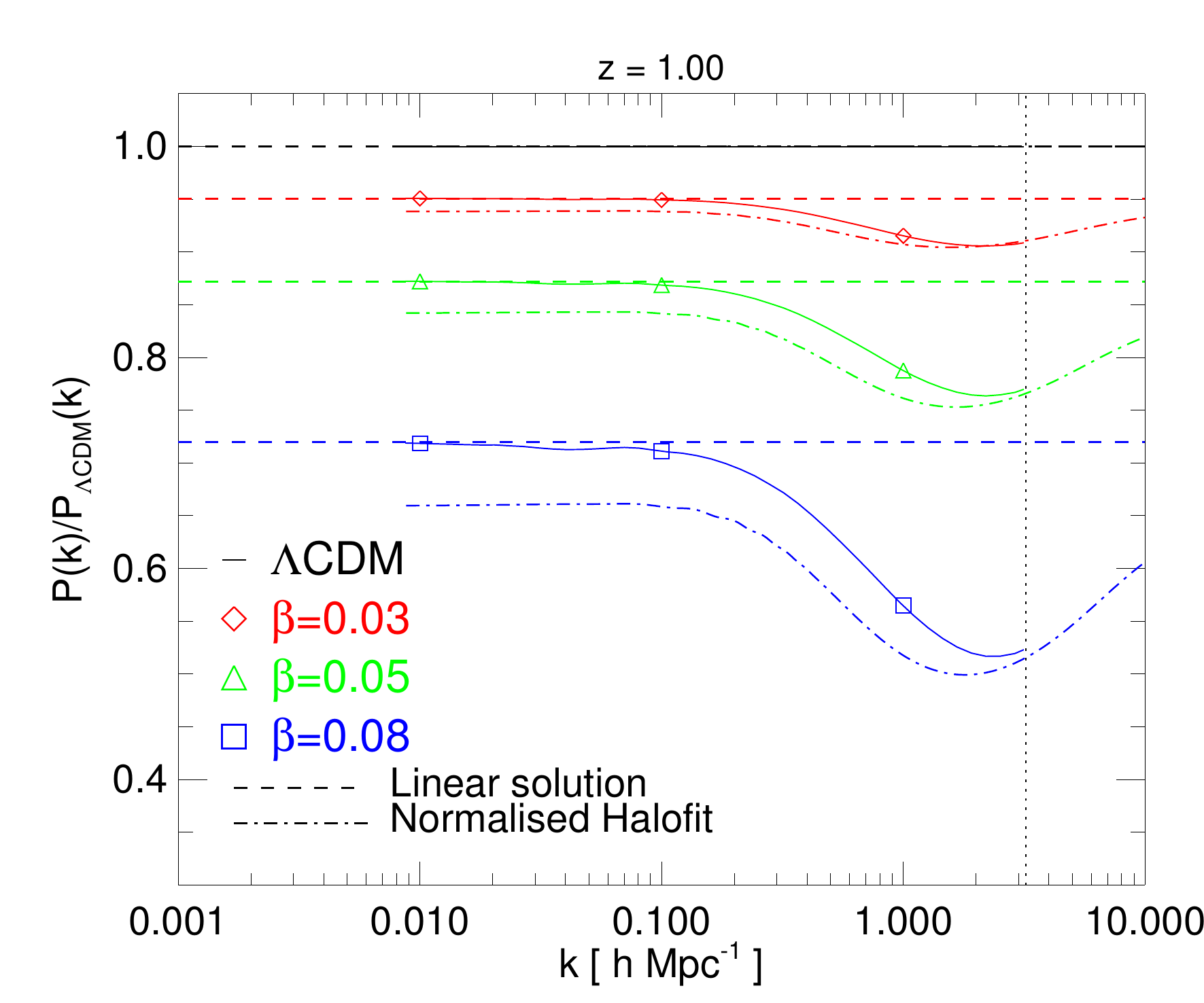}
\caption{The power spectrum ratio with respect to $\Lambda$CDM at $z=0$ ({\em left}) and $z=1$ ({\em right}) for all the CIDER cosmologies under investigation. The dashed line represents the expected linear ratio based on the solution of the perturbations equations, while the dot-dashed curve is the power ratio for a $\Lambda $CDM cosmology with the same $\sigma_{8}(z=0)$ as each of the CIDER models, as computed by {\small CAMB+HaloFit}.}
\label{fig:power_spectrum_ratio}
\end{figure*}

First of all, we compare the large-scale matter distribution in terms of the non-linear matter power spectrum $P(k,z)$ extracted from the simulations. In Fig.~\ref{fig:power_spectrum_ratio} we plot the ratio of the power spectra in the different CIDER models to the reference $\Lambda $CDM cosmology at $z=0$ (left panel) and $z=1$ (right panel). The vertical dotted line represents the Nyquist frequency of the primary grid used to compute the spectra, corresponding to $k_{\rm Ny} \equiv \pi N/L\approx 3.2\, h/$Mpc.

The horizontal dashed lines represent the expected linear ratio of the power spectra based on the the linear growth factor obtained as the numerical solution of Eqs.~(\ref{pertc}-\ref{pertb}) for each model. As one can see from the figures, the simulated spectra perfectly track the expected linear ratio at scales $k\lesssim 0.1\, h/$Mpc. 

The dot-dashed curves correspond to the expected non-linear power spectrum ratio to the reference cosmology as obtained from the public Boltzmann code  {\small CAMB} \citep[][]{camb} with the {\small HaloFit} non-linear module \citep[][]{Smith_etal_2003,Takahashi_etal_2012} for a set of $\Lambda $CDM models having the same value of $\sigma _{8}(z=0)$ as each of the CIDER cosmologies under investigation. As the figure shows, at $z=0$ the CAMB+HaloFit ratios converge to the expected linear ratios -- thereby matching with the simulation predictions as well -- at large scales, while featuring a significant suppression of power at smaller scales due to a delayed development of non-linearities as a consequence of the lower linear normalisation. Nonetheless, the simulation results (solid curves) exhibit an even stronger suppression at small scales compared with this {\small CAMB+HaloFit} prediction, showing the imprints of the non-linear effects of the DE-CDM interaction.
At $z=1$, the offset between the CAMB+HaloFit ratios and the linear expectation at large scales shows the different linear growth rate between the CIDER models and a standard $\Lambda $CDM cosmology with the same $\sigma _{8}(z=0)$.

This comparison is particularly instructive as it allows to highlight the distinct footprints of the CIDER non-linear evolution with respect to a standard $\Lambda $CDM cosmology with the same background expansion and the same linear perturbations normalisation, namely: {\em (i)} a different redshift evolution of the linear growth amplitude and {\em (ii)} a further suppression of power at the smallest and most non-linear scales driven by the  friction term of Eqs~(\ref{pertb}-\ref{pertc}) that counteracts the gravitational collapse of bound structures. The latter effect, characterising the fully non-linear regime of structure formation, may allow to distinguish the CIDER cosmology from a standard $\Lambda $CDM  model through measurements at a single redshift. As one can see from the plots, both the linear-scales deviation from a $\Lambda $CDM growth and the non-linear scale-dependent suppression of power are proportional to the strength of the coupling, ranging (at $z=0$) from a few percent effect for the weakest coupling ($\beta = 0.03$) to a $\sim 35\%$ effect for the strongest coupling ($\beta = 0.08$) under investigation.

\subsubsection{The baryon-DM gravitational bias}
\begin{figure*}

\includegraphics[width=0.49\textwidth]{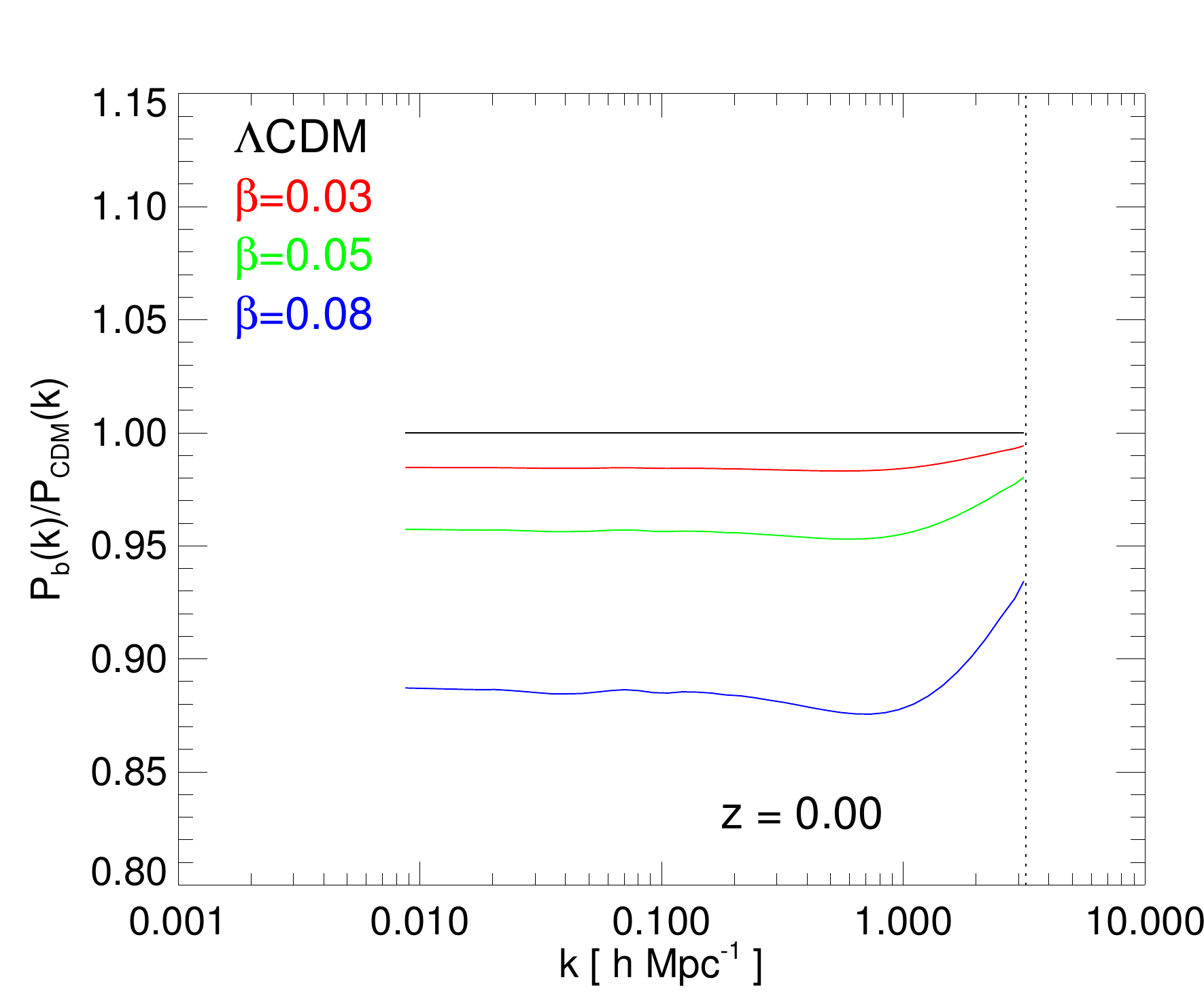}
\includegraphics[width=0.49\textwidth]{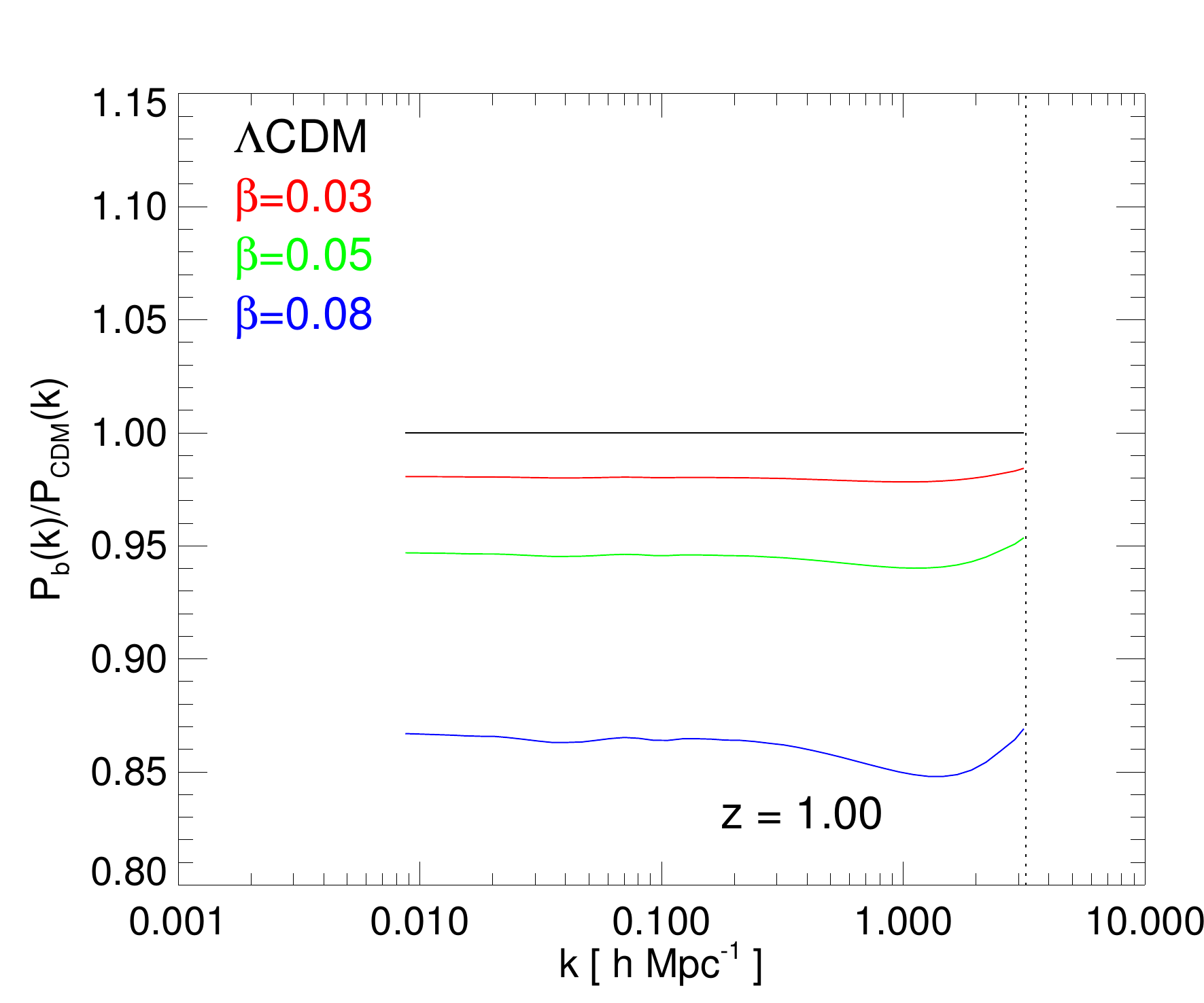}
\caption{The ratio of the baryon and CDM power spectra with respect to the standard $\Lambda$CDM case at $z=0$ ({\em left}) and $z=1$ ({\em right}). The violation of the Weak Equivalence Principle induces a different evolution of the baryon and CDM density perturbations as compared to the standard $\Lambda $CDM cosmology, resulting in a scale-independent suppression of the baryonic perturbations amplitude at large linear scales, which acquires a clear scale dependence at smaller scales due to to he onset of non-linear evolution.}
\label{fig:gravitational_bias}
\end{figure*}

One of the most distinctive features of interacting Dark Energy models in general is the violation of the Weak Equivalence Principle, i.e. the fact that different matter species experience different gravitational accelerations arising from the same global matter distribution. This applies also to the CIDER scenario, as discussed above in Section~\ref{sec:linear}. Therefore, in these cosmological models we do expect that Dark Matter and baryons may develop some bias in their spatial distribution as a consequence of their different effective gravitational accelerations, on top of the hydrodynamical bias developing at small scales due to the pressure forces acting on baryons and to radiative effects associated with star and galaxy formation, which are anyway not included in our numerical treatment. In this respect, our simulations include the baryonic matter fraction as an additional family of collisionless particles,  -- i.e. we switch off the SPH hydrodynamic forces on baryons and any other astrophysical effect -- thereby allowing us to observe the impact of the Weak Equivalence Principle violation without any additional source of possible offsets between the baryons and CDM distributions. 

The simplest statistics to quantify this effect is the so called gravitational bias, defined as the ratio between the baryon and the CDM power spectra, as a function of scale, $P_{b}(k)/P_{c}(k)$. Due to the additional accelerations experienced by CDM particles as a consequence of both the fifth-force and  the extra friction terms, we  expect CDM perturbations to evolve faster than baryon perturbations in the linear regime, while in the non-linear regime the onset of virialisation may lead to a more complex interplay of these two additional forces \citep[][]{Baldi_2011b}. In Fig.~\ref{fig:gravitational_bias} we plot the gravitational bias with respect to the $\Lambda $CDM reference cosmology at $z=0$ and $z=1$ for all the CIDER models under investigation. As the figures show, the CIDER models feature a progressively stronger and scale-independent suppression of the gravitational bias for increasing values of the coupling at large, linear scales, with a clear tendency to reduce the effect at the smallest scales available in our power spectra due to the onset of nonlinear evolution where the friction term slows down the collapse of the CDM component (as discussed above). 

\subsection{Statistical and structural properties of Dark Matter haloes}
\label{sec:halos}

\subsubsection{Halo Mass Function}
\begin{figure*}
\includegraphics[width=0.49\textwidth]{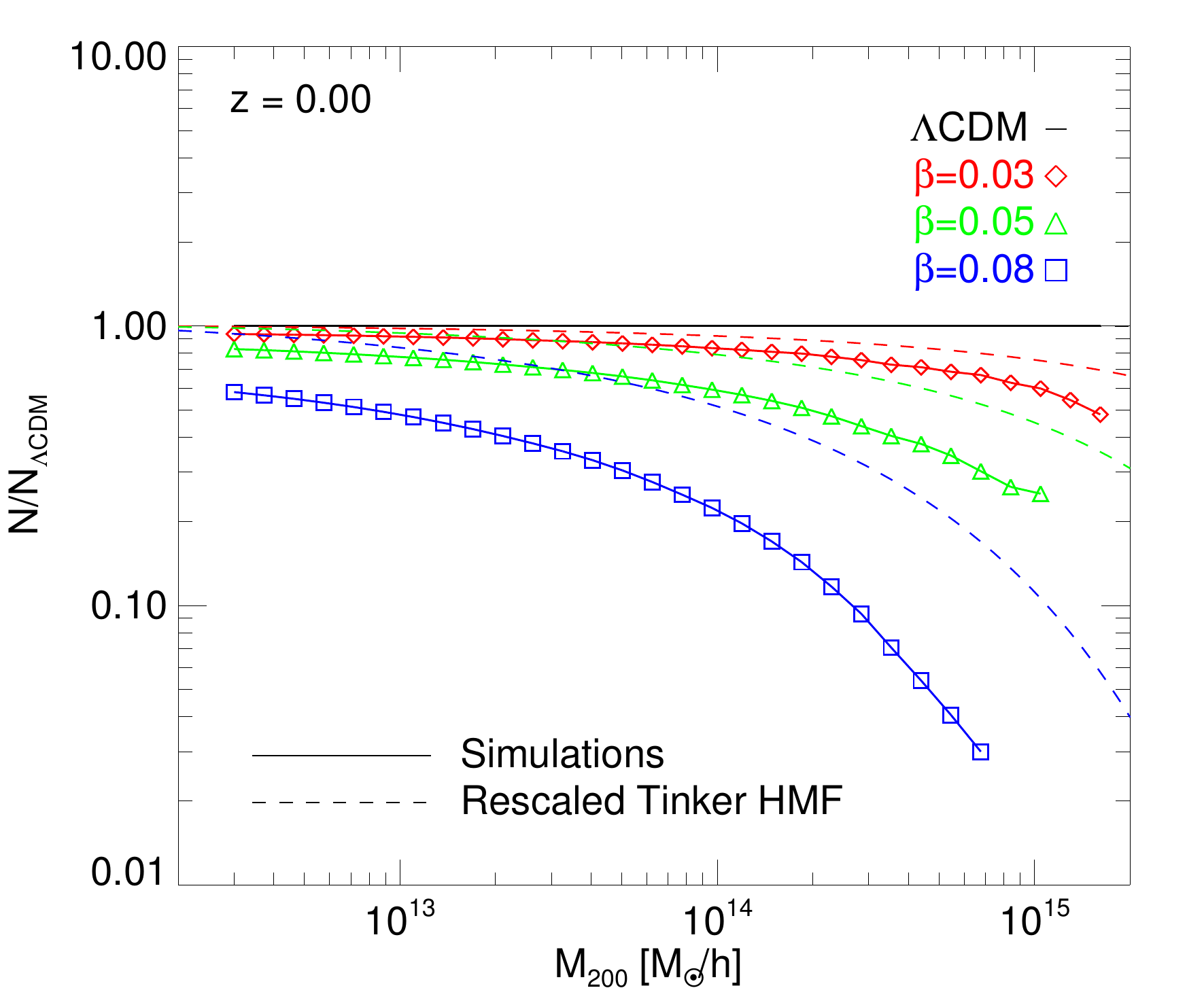}
\includegraphics[width=0.49\textwidth]{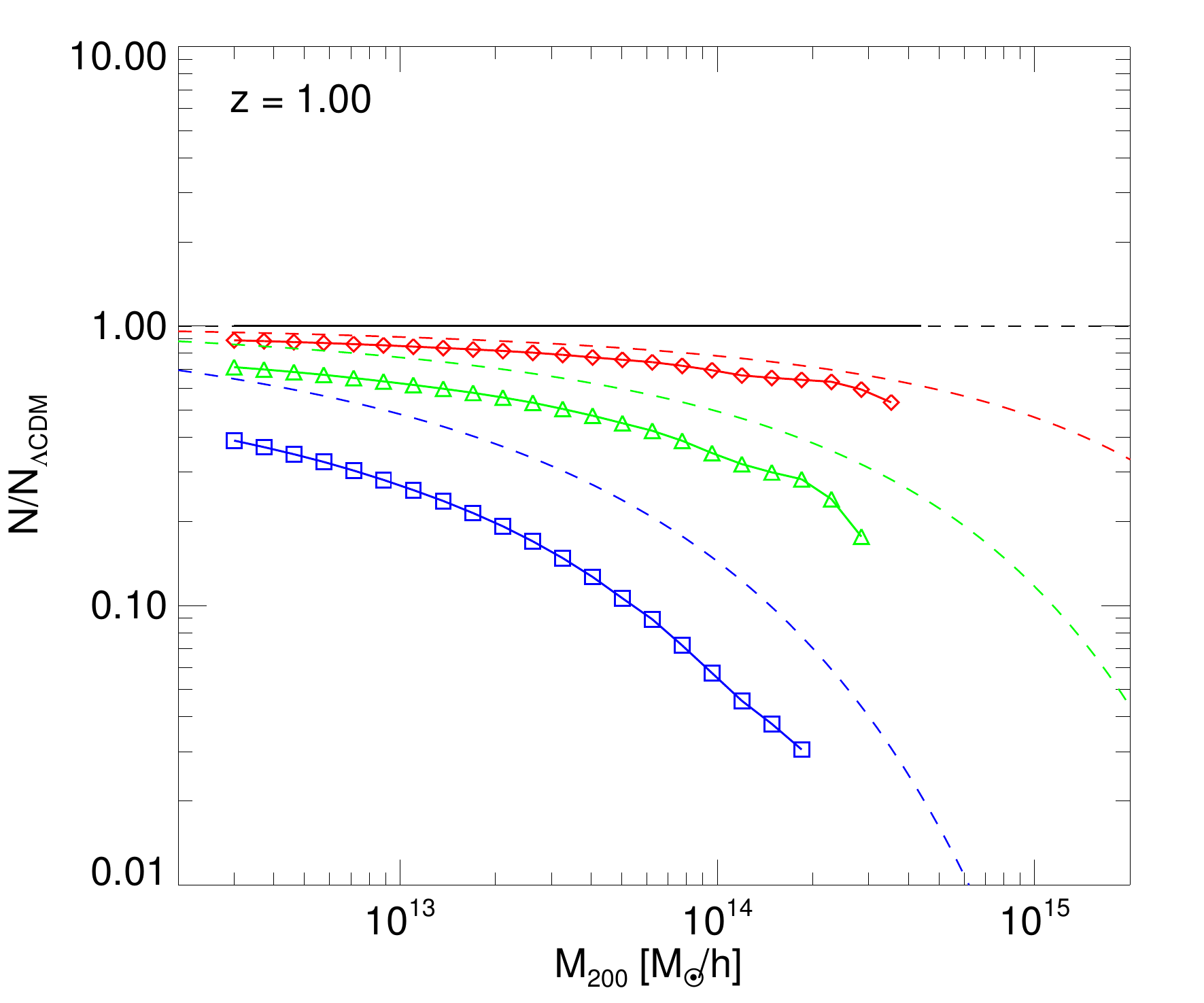}
\caption{The Halo Mass Function ratio with respect to $\Lambda$CDM at $z=0$ ({\em left}) and $z=1$ ({\em right}). Solid curves with symbols indicate the measured mass function ratio from the simulations, while dashed curves show the ratio predicted by the theoretical \citet{Tinker_etal_2010} Halo Mass Function for a $\Lambda $CDM cosmology with the same $\sigma_{8}(z=0)$ normalisation as the various CIDER realisations. The non-linear dynamics of the CIDER models determines an additional suppression of the halo abundance compared to a $\Lambda $CDM cosmology with identical background expansion history and linear power normalisation.}
\label{fig:mass_function_ratio}
\end{figure*}

We now move to investigate the statistical properties of collapsed (i.e. gravitationally bound) structures in our simulations, starting from the Halo Mass Function. For all our FoF halos we have computed the spherical overdensity mass $M_{200}$ and radius $R_{200}$ at an overdensity threshold of $200$ times with respect to both the critical density and the mean density of the universe around the most bound particle of the main substructure identified by {\small SUBFIND}. With these we have computed the cumulative halo mass function of the sample. In Fig.~\ref{fig:mass_function_ratio} we display as solid curves the ratio of the $M_{200,mean}$ Halo Mass Functions of all the CIDER models with respect to the standard $\Lambda $CDM cosmology, at $z=0$ and $z=1$. Both plots show a significant suppression of the abundance of halos for increasing values of the coupling $\beta $, with a clear  mass dependence making the effect more pronounced at the largest masses in the sample. The relative suppression reaches a value of $\approx 96\%$ at the largest masses for the strongest value of the coupling at $z=0$, so that the most massive halos identified in the $\Lambda $CDM cosmology are completely absent in the $\beta =0.08$ model. As a comparison, we have computed the Tinker Halo Mass Function \citep[][]{Tinker_etal_2010} for a set of $\Lambda $CDM cosmologies with the same cosmological parameters as our simulations but for different values of $\sigma _{8}(z=0)$ corresponding to the values obtained in the three CIDER cosmologies, and we show the ratio of these theoretical mass functions to the reference $\Lambda $CDM model as dashed curves in both panels of Fig.~\ref{fig:mass_function_ratio}. As it clearly appears from the figures, the actual suppression of the Halo Mass Function in the simulations greatly exceeds the one predicted for a $\Lambda $CDM cosmology with the same $\sigma_{8}$ value. This is consistent with what found above on the matter power spectrum, where the smallest scales show an additional suppression of power with respect to a $\Lambda $CDM model with the same linear amplitude normalisation. We notice that this effect also shows a weak redshift dependence, with the abundance at a fixed halo mass being more strongly suppressed at higher redshifts.

\subsubsection{Halo density profiles}
\label{sec:halo_profiles}

\begin{figure*}
\includegraphics[width=\textwidth]{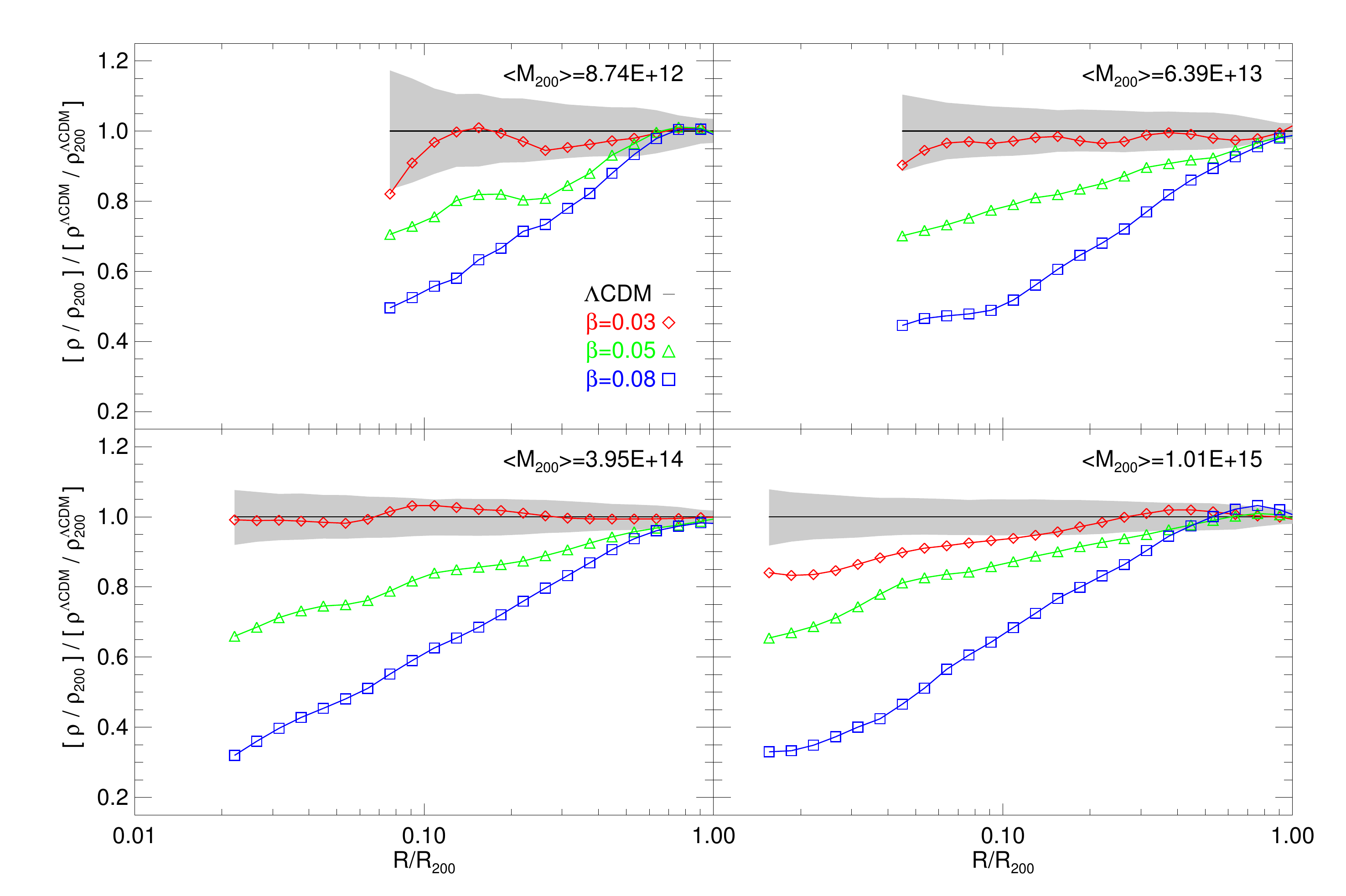}
\caption{The stacked halo density profile ratio with respect to $\Lambda$CDM in four mass bins at $z=0$ for all the models under investigation. Radial density profiles have been computed in a set of 30  logarithmically-equispaced radial bins in units of the halo radius $R_{200,crit}$, and the plotted profile ratios are truncated at the bin where the median $R_{200,crit}$ in the $\Lambda $CDM simulation of each mass bin is smaller than the gravitational softening of the simulation $e_{g}=25$ kpc$/h$. The grey shaded area represents a $2−\sigma$ uncertainty on the stacked profile based on a bootstrap resampling procedure.}
\label{fig:halo_profiles}
\end{figure*}

For all our simulations, we divided the halo sample at $z=0$ into four mass bins corresponding to values of the $M_{200,crit}$ spherical overdensity mass in the ranges $ \left[ 0-10^{13}\right] $,$\left[ 10^{13}-7.5\times 10^{13}\right] $,$\left[ 7.5\times 10^{13}-5\times 10^{14}\right] $, $\left[ 5\times 10^{14}-\infty \right] $, and for $100$ randomly selected halos in each of these mass bins we computed spherically averaged density profiles in $30$ logarithmically equispaced radial bins centered around the halo most bound particle. In Fig.~\ref{fig:halo_profiles} we show the ratio of the stacked halo density profiles in each bin with respect to the reference $\Lambda $CDM model, for all our CIDER cosmologies. As the plot very clearly shows, the density profiles feature a progressively stronger suppression for increasing values of the coupling, in all mass bins. The grey-shaded areas indicate the $2-\sigma $ statistical confidence on the $\Lambda $CDM profiles based on a bootstrap resampling of the $100$ randomly selected halos in each mass bin. Therefore, while the weakest coupling $\beta = 0.03$ provides a significant suppression only for the most massive halos, the $\beta = 0.05$ and $\beta = 0.08$ models show a statistically significant deviation from the $\Lambda $CDM expectation at all masses. This is a well known effect of interacting dark energy models with constant couplings, but it is the first time that this is observed in the context of the CIDER models, where the background effects of the coupling are absent by construction.

\subsubsection{Concentrations}
\begin{figure}

\includegraphics[width=\columnwidth]{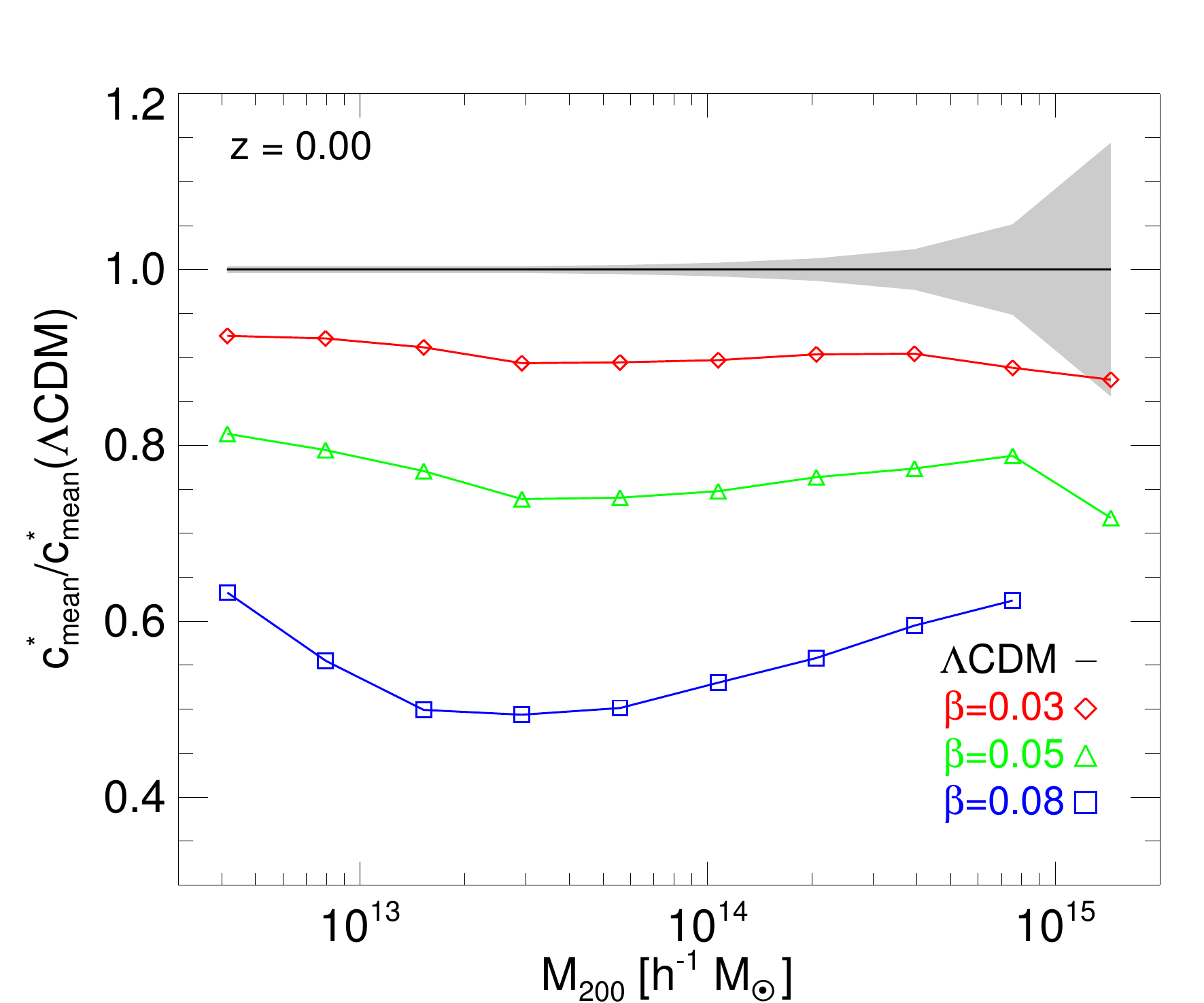}
\caption{The ratio with respect to $\Lambda$CDM of the concentration-mass relation at $z=0$. The grey-shaded region represents the Poissonian error on the ratio for the $\Lambda $CDM cosmology.}
\label{fig:c_vmax_ratio}
\end{figure}

As a complement to the stacked density profile ratios discussed above, we have also computed for all halos in our sample (i.e. without any random selection) the halo concentration $c^{\star }$ defined as \citep[][]{Aquarius}:
\begin{equation}
\label{conc}
\frac{200}{3}\frac{c^{\star 3}}{\ln (1+c^{\star }) - c^{\star }/(1+c^{\star })} = 7.213~\delta _{V}
\end{equation}
with $\delta _{V}$ defined as:
\begin{equation}
\delta _{V} = 2\left( \frac{V_{\rm max}}{H_{0}r_{\rm max}}\right) ^{2}
\end{equation}
where $V_{\rm max}$ and $r_{\rm max}$ are the maximum rotational velocity of the halo and the radius at which this velocity peak is located, respectively. 
In Fig.~\ref{fig:c_vmax_ratio} we display the evolution of the concentration-mass relation with coupling in the CIDER cosmologies by plotting the halo concentrations computed according to Eq~(\ref{conc}) as a function of the spherical overdensity mass $M_{200,crit}$, in units of the reference $\Lambda $CDM cosmology. All models show a clear suppression of the halo concentrations over the whole mass range of our halo sample, with the suppression increasing with increasing coupling. The grey-shaded area represents the Poissonian error on the ratio for the $\Lambda $CDM halo sample. Again, the observed suppression of halo concentration is a well known feature of interacting Dark Energy models with constant couplings, but we highlight it here for the first time for models with a background expansion history indistinguishable from $\Lambda $CDM. Therefore, for the CIDER cosmologies it is possible to obtain a quantitative assessment of the concentration suppression due to the modified gravitational dynamics alone, without possible additional effects arising from a different expansion history.

\subsection{Statistical and structural properties of cosmic voids}

As a final test of the impact of the CIDER models on structure formation we investigate the properties of the cosmic voids identified in our simulations through the procedure detailed in Section~\ref{sec:voidfinding}. 

\subsubsection{Abundance of cosmic voids}
\label{sec:voids_abundance}
\begin{figure*}

\includegraphics[width=\textwidth]{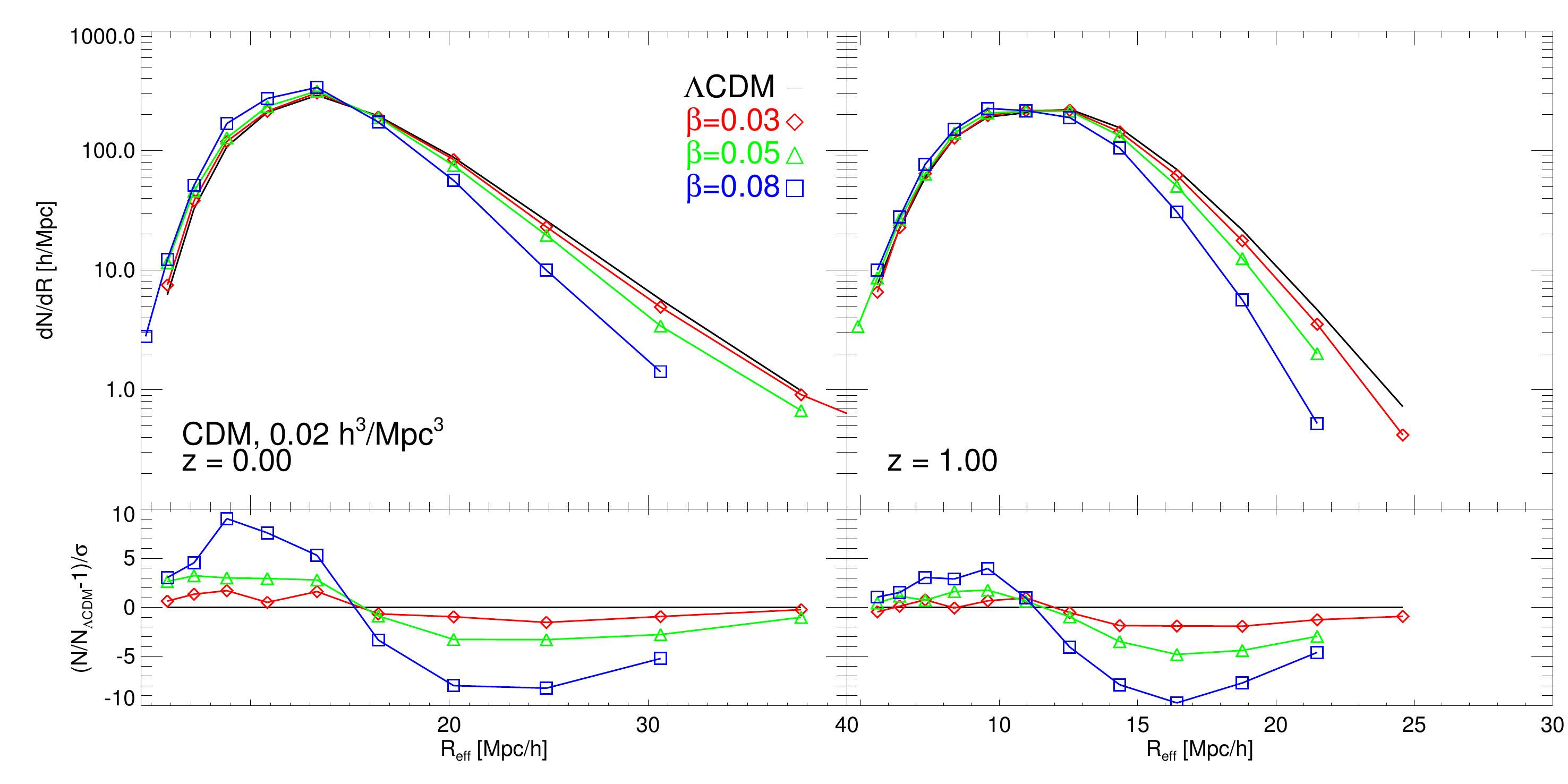}
\caption{The differential void size function at $z=0$ and $z=1$ based on the CDM voids catalogues extracted from a random subsampling of the CDM particle distribution for all the cosmological models under investigation. Bottom panles show the relative difference with respect to the reference $\Lambda $CDM cosmology in units of the statistical significance of the ratio.}
\label{fig:voids_abundance}
\end{figure*}

We start by computing the abundance of cosmic voids as a function of their effective radius $R_{\rm eff}$ as defined in Equation~(\ref{effective_radius}). This is known to be a powerful probe for cosmological parameter estimation \citep[][]{Ronconi_etal_2019,Contarini_etal_2019}, especially when joint with other complementary probes \citep[][]{Hamaus_etal_2020,Hamaus_etal_2021}, and to constrain extensions to the standard cosmological scenario \citep[see e.g.][]{Cai_Padilla_Li_2015,Pisani_etal_2015,Pollina_etal_2016,Voivodic_etal_2017,Verza_etal_2019,Schuster_etal_2019,Contarini_etal_2020}. 

In Fig.~\ref{fig:voids_abundance} we display the differential void size function $dN/dR_{\rm eff}$ as a function of the void effective radius $R_{\rm eff}$ for the reference $\Lambda $CDM cosmology (solid black) and for the three CIDER models investigated in the present work, at $z=0$ and $z=1$. The void size functions show their typical shape characterised by a growing abundance at small radii (primarily due to the void radii becoming comparable with the spatial resolution of the downsampled matter distribution, thereby making smaller voids undetectable) reaching a peak around $10-15$ Mpc$/h$ followed by an exponential decrease at larger radii, indicating the hierarchical build up of larger voids from the merging of smaller ones. 

The comparison among the different models shows very clearly the effect of the CIDER cosmologies on the size function, resulting in a suppressed abundance of the larger voids (i.e. on the right of the peak) and a corresponding increase of the smaller ones (i.e. on the left of the peak), witnessing a delay in the hierarchical assembly of voids due to the suppression of the growth rate of cosmic structures. In the lower panels of Fig.~\ref{fig:voids_abundance} we show the relative deviation  of the void size functions  to the reference $\Lambda $CDM case $N/N_{\Lambda {\rm CDM}}$, in units of the
statistical significance $\sigma $ computed by propagating the Poisson
noise in each bin of effective radius to the relative difference. The effect is roughly proportional to the value of the coupling $\beta$, and reaches a significance of $7-10\, \sigma $ for the most extreme CIDER scenario. This result shows how voids abundance could be another powerful probe to constrain this class of interacting DE scenarios.

\subsubsection{Void density profiles}
\begin{figure*}
\includegraphics[width=\textwidth]{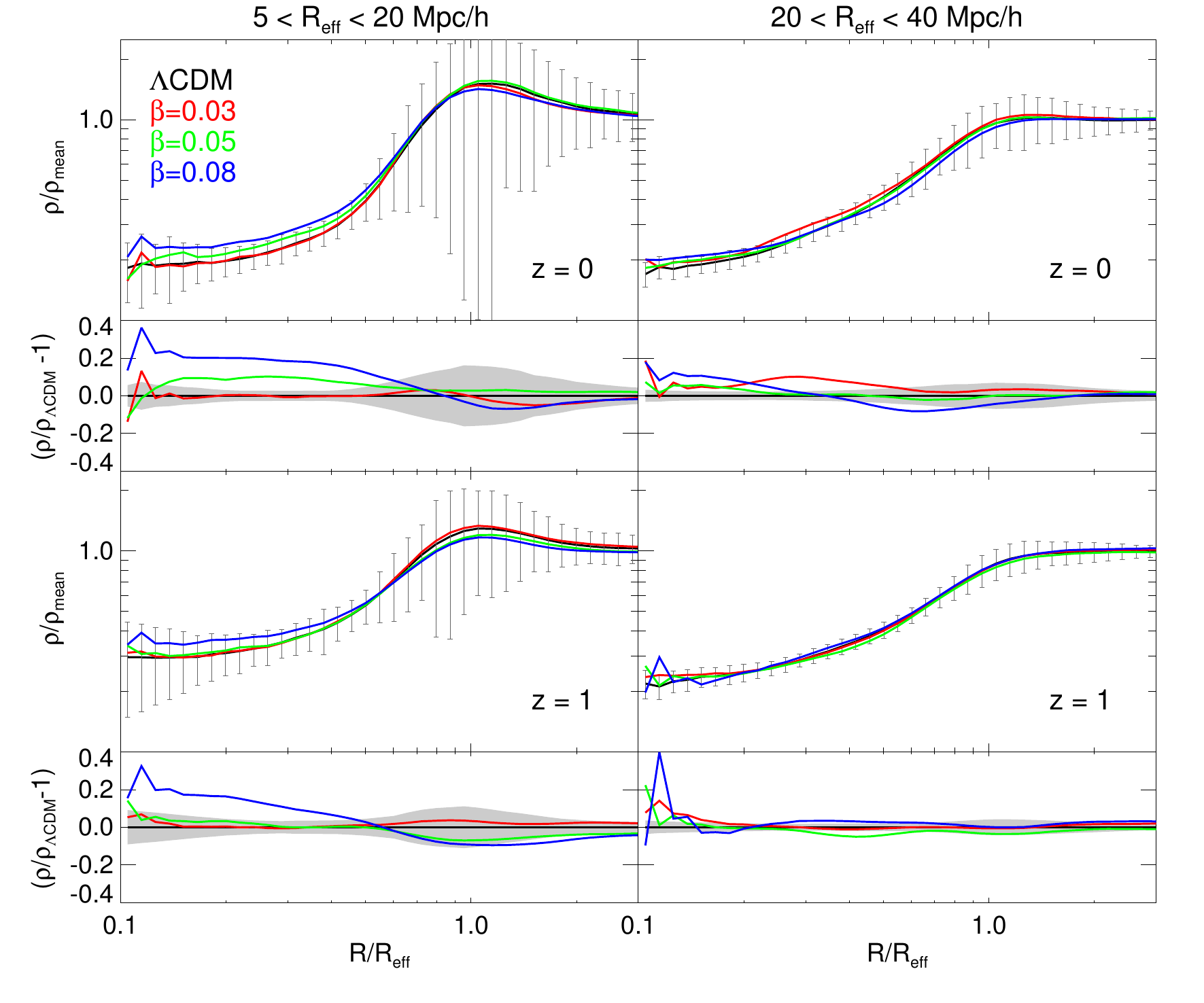}
\caption{The stacked void density profiles in two different ranges of void effective radius at $z=0$ ({\em top}) and $z=1$ ({\em bottom}). In each plot, the upper panels show the spherically averaged stacked profiles for all the cosmological models under investigation while the lower panels display the ratio to the reference $\Lambda $CDM scenario. Error bars represent the statistical Possonian errors on the average density in each radial bin as computed from the $\Lambda $CDM simulation, while the grey-shaded area shows the statistical significance of the ratio computed through a {\em bootstrap} resampling procedure.}
\label{fig:voids_profiles}
\end{figure*}

We finally investigate the structural properties of cosmic voids
in the various CIDER models by computing the stacked radial density profiles around the voids centers identified in the $\Lambda $CDM simulation for a subsample of 100 randomly selected voids
within two bins of void effective radius, namely $R_{\rm eff} \in  \left\{5-20\,, 20-40\right\}$ Mpc$/h$. We display the outcomes of such comparison in Fig.~\ref{fig:voids_profiles} for these two $R_{\rm eff}$ bins (left and right plots, respectively) at $z=0$ and $z=1$ (upper and lower plots, respectively). In each plot, the top panel presents the density profile of the different models, where the error bars represent the Poissonian error on the mean density value in each of the $50$ logarithmically equispaced radial bins used to build the profiles, based on the number of voids considered in the $\Lambda $CDM stacking; the bottom panel shows the relative deviation from the reference $\Lambda $CDM model, where the grey-shaded area indicates the $1-\sigma $ statistical significance according to a bootstrap estimation. The comparison shows that cosmic voids in CIDER cosmologies are shallower than their $\Lambda $CDM counterparts, with a less pronounced compensation wall, with the effect being directly proportional to the coupling constant $\beta $ and reaching a deviation of $30-40\%$ for the most extreme value of $\beta = 0.08$ in the inner regions of the smaller voids, while the larger voids show milder and less significant deviations. Therefore, the CIDER cosmology results in less empty voids which may determine a lower amplitude of the weak lensing signal at small scales. This feature also points towards a possible alleviation of the $\sigma _{8}$ tension.

\section{Conclusions}
\label{sec:conclusions}

Cosmological models beyond the standard $\Lambda $CDM scenario have been proposed and investigated over the past decades based on a wide range of theoretical and observational motivations, and still represent a highly active field of research in view of the  wealth of high-precision data expected from the upcoming era of large-volume cosmological surveys. In particular, the search for alternative models has gained relevance after the emergence of the two well-known observational tensions regarding the value of the local Hubble constant $H_{0}$ and the amplitude of density fluctuations $\sigma _{8}$ at low redshifts. 

Although both issues may be resulting from unaccounted systematics or data interpretation problems, none of the many and thorough  investigations in this direction have so far succeeded in reducing these tensions below the level of statistical significance.

On the other hand, also extended cosmological models are still struggling to provide satisfactory solutions to both problems at the same time, allowing in most cases to address only one of the aforementioned tensions. In particular, in the context of Dark Energy and Modified Gravity scenarios, it is generally difficult to address the $\sigma _{8}$ tension as most of such models predict an additional attractive force on top of standard gravity, thereby giving rise to an enhanced growth of perturbations and thus a higher (and not lower) amplitude of linear density fluctuations at low redshifts. Therefore, it is interesting to investigate thoroughly those (few) models that may show the appealing feature of suppressing structure growth at late times. This is the case e.g. for Dark Scattering scenarios \citep[see e.g.][]{Simpson_2010,Baldi_Simpson_2017}, {\em Bouncing} interacting Dark Energy \citep[][]{Baldi_2011b} or some particular realisations of scalar-tensor Modified Gravity theories \citep[][]{Wittner_etal_2020}. 

In a recent work, \citet{Barros_etal_2019} have proposed a particular class of coupled Dark Energy \citep[][]{Amendola_2000,Amendola_2004} models that may result in a slower growth of structures as compared to $\Lambda $CDM. The most remarkable feature of such models, that we termed here as the {\em Constrained Interacting Dark Energy} (or CIDER) cosmologies, is that the background expansion history is constrained to be identical to a reference $\Lambda $CDM model by construction, similarly to what happens -- in a different context -- for the \citet{Hu_Sawicki_2007} realisation of $f(R)$ Modified Gravity. Therefore, although obviously not allowing to ease the $H_{0}$ tension (at least in their original formulation) these CIDER cosmologies result in a suppression of structure formation at late times thereby possibly accommodating low-redshift observations of the density perturbations amplitude, without further exacerbating the $H_{0}$ tension.

While some insight into the non-linear evolution of the CIDER models has been provided by studying the spherical collapse process in \citet{Barros_etal_2020}, no extended analysis of non-linear structure formation has been performed for these scenarios so far, and we have provided in the present work the first extensive investigation in this direction by presenting a suite of full cosmological N-body simulations implementing the physics of the CIDER  models and discussing their basic outcomes. More detailed analysis of these simulations for specific cosmological observables will be discussed in future works.

More specifically, we have run a set of four cosmological simulations for one reference $\Lambda $CDM model and three realisations of the CIDER scenario for different values of the coupling constant $\beta$. The simulations feature a periodic box of $1$ Gpc$/h$ per side with $2\times 1024^3$ particles -- including both baryons and CDM species -- in order to properly implement the violation of the Weak Equivalence Principle that characterises all interacting Dark Energy models, including the CIDER scenario. For all these simulations, we have produced density power spectra for both particle species, halo and subhalo catalogues, and voids identification, and we have analysed these data with respect to several possible observables. These simulations are part of a more extended initiative -- the {\small CoDECS2} project -- that is aimed at providing a follow-up of the {\small CoDECS} project \citep[][]{CoDECS} by updating the landscape of simulated interacting DE cosmologies with some of the relevant (and still viable) models proposed in the last decade. In this work, we have thus discussed the results obtained for the first class of such models -- the CIDER scenario -- covered by the {\small CoDECS2} project.\\

In particular, our main findings can be summarised as follows:
\begin{itemize}
    \item [$\star $] The large-scale density distribution (Fig.~\ref{fig:mass_maps}) shows the same topology and shape of the cosmic web (as expected due to the identical initial conditions) for all our simulations; still, the visual comparison between the $\Lambda $CDM and the most extreme CIDER model with $\beta =0.08$ already allows to observe differences in the location and height of the density peaks associated with individual halos, showing a clearly less evolved distribution for the CIDER universe;\\
    \item [$\star $] The non-linear matter power spectrum (Fig.~\ref{fig:power_spectrum_ratio}) shows a clear scale-independent suppression at large scales, with the effect being directly proportional to the strength of the coupling constant $\beta$; at smaller scales, a further scale-dependent suppression occurs, which is primarily due to the shift in the non-linearity scale associated with the lower linear amplitude; nonetheless, by comparing the spectra of each CIDER model with a non-linear spectrum of a $\Lambda $CDM cosmology calibrated to have the same linear amplitude (i.e. the same $\sigma _{8}$, thereby cancelling any effect due to the difference in linear normalisation), the CIDER models still show a further small-scale suppression which appears to be related to the effect of the velocity-dependent {\em drag} acceleration characterising these models; such extra suppression may be crucial to break degeneracies with $\sigma _{8}$ through e.g. weak lensing observations at large multipoles, as shown for other interacting DE models by e.g. \citet{Pace_etal_2015,Giocoli_etal_2015};\\
    \item [$\star $] The Weak Equivalence Principle violation appears clearly in the ratio between the baryon and CDM non-linear matter power spectra (Fig.~\ref{fig:gravitational_bias}), which shows a suppression directly proportional to the coupling strength $\beta $; this is due to the fact that, besides the overall slower growth due to the progressive decrease of the dark matter particle mass (that affects both species through the total gravitational potential), dark matter particles also experience an extra acceleration due to the combination of the fifth-force and the velocity-dependent drag terms; this determines a lag of the baryonic overdensity amplitudes with respect to the dark matter ones;\\
    \item [$\star $] The abundance of halos quantified by the Halo Mass Function (Fig.~\ref{fig:mass_function_ratio}) shows a suppression over the whole mass range covered by our halo catalogue, with a clear exponential mass dependence, and with an amplitude proportional to the coupling strength $\beta $; also in this case, the effect significantly exceeds what would be expected for a $\Lambda $CDM cosmology with the same linear perturbations normalisation at $z=0$, as computed through e.g. the \citet{Tinker_etal_2010} fit to the Halo Mass Function; therefore, the non-linear evolution of the model plays a crucial role in the buildup of collapsed structures and may also allow to break degeneracies with standard cosmological parameters;\\ 
    \item [$\star $] The importance of highly non-linear effects in shaping the structural properties of halos is also very clearly highlighted by the comparison of stacked halo density profiles (Fig.~\ref{fig:halo_profiles}) showing a clear suppression of the inner overdensity of halos at all masses; while the suppression may be weak and barely significant for the smallest coupling value under investigation ($\beta = 0.03$) it is instead quite relevant for the larger coupling values reaching $\approx 60\%$ for $\beta = 0.08$;\\
    \item [$\star $] As a consequence of the overdensity suppression in the inner regions of collapsed objects, also the concentration-Mass relation (Fig.~\ref{fig:c_vmax_ratio}) is suppressed at all masses in the CIDER models with respect to $\Lambda $CDM, with an amplitude proportional to the coupling constant, ranging from $\approx 5-10\%$ to $\approx 40-50\%$ for the couplings considered in this work;\\
    \item [$\star $] Cosmic voids are also affected by the slower growth of density perturbations, with a significantly lower abundance (Fig.~\ref{fig:voids_abundance}) of the largest voids ($\gtrsim 10$ Mpc$/h$) and a corresponding increased abundance of the smaller ones, witnessing a delay in the process of hierarchical merging of cosmic voids; also in this case the deviation is directly proportional to the strength of the coupling constant $\beta $;\\
    \item [$\star $] Finally, we investigated the stacked density profiles of cosmic voids (Fig.~\ref{fig:voids_profiles}) showing how the CIDER cosmologies tend to have shallower voids with less underdense inner regions and a less pronounced compensation wall at their boundary; such feature also goes in the direction of a weaker lensing power amplitude at the corresponding angular scales thereby possibly alleviating the $\sigma _{8}$ tension.
\end{itemize}

To conclude, we have performed for the first time cosmological simulations of an interacting Dark Energy scenario constructed by fixing the background expansion history to that of the standard $\Lambda $CDM model, and presented their main outcomes in terms of a set of basic cosmological observables showing distinctive footprints of the effects of the Dark Energy interaction on the properties of large-scale structures. Our results confirm the tendency of these particular cosmological models to suppress structure formation, thereby providing a possible avenue to address the persisting observational $\sigma _{8}$ tension.

\section*{Acknowledgments}

I am deeply thankful to Luca Amendola and Nelson Nunes for reading the draft and providing useful comments.
This work has been supported by the project ``Combining Cosmic Microwave Background and Large Scale Structure data: an Integrated Approach for Addressing Fundamental Questions in Cosmology", funded by the MIUR Progetti di Ricerca di Rilevante Interesse Nazionale (PRIN) Bando 2017 - grant 2017YJYZAH. The simulations have been run on the Marconi supercomputing cluster at Cineca, thanks to the INFN-Euclid allocated budget, and the data analysis has been performed on the parallel computing cluster of the Open Physics Hub (\url{https://site.unibo.it/openphysicshub/en}) at the Physics and Astronomy Department in Bologna.

\bibliographystyle{mnras}
\bibliography{CIDER}

\begin{thebibliography}{}
\makeatletter
\relax
\def\mn@urlcharsother{\let\do\@makeother \do\$\do\&\do\#\do\^\do\_\do\%\do\~}
\def\mn@doi{\begingroup\mn@urlcharsother \@ifnextchar [ {\mn@doi@}
  {\mn@doi@[]}}
\def\mn@doi@[#1]#2{\def\@tempa{#1}\ifx\@tempa\@empty \href
  {http://dx.doi.org/#2} {doi:#2}\else \href {http://dx.doi.org/#2} {#1}\fi
  \endgroup}
\def\mn@eprint#1#2{\mn@eprint@#1:#2::\@nil}
\def\mn@eprint@arXiv#1{\href {http://arxiv.org/abs/#1} {{\tt arXiv:#1}}}
\def\mn@eprint@dblp#1{\href {http://dblp.uni-trier.de/rec/bibtex/#1.xml}
  {dblp:#1}}
\def\mn@eprint@#1:#2:#3:#4\@nil{\def\@tempa {#1}\def\@tempb {#2}\def\@tempc
  {#3}\ifx \@tempc \@empty \let \@tempc \@tempb \let \@tempb \@tempa \fi \ifx
  \@tempb \@empty \def\@tempb {arXiv}\fi \@ifundefined
  {mn@eprint@\@tempb}{\@tempb:\@tempc}{\expandafter \expandafter \csname
  mn@eprint@\@tempb\endcsname \expandafter{\@tempc}}}

\bibitem[\protect\citeauthoryear{{Abbott} et~al.,}{{Abbott}
  et~al.}{2018}]{Abbott_etal_2018}
{Abbott} T.~M.~C.,  et~al., 2018, \mn@doi [\prd] {10.1103/PhysRevD.98.043526},
  \href {https://ui.adsabs.harvard.edu/abs/2018PhRvD..98d3526A} {98, 043526}

\bibitem[\protect\citeauthoryear{Abbott et~al.}{Abbott
  et~al.}{2020a}]{Abbott_etal_2020}
Abbott T. M.~C.,  et~al., 2020a, \mn@doi [Phys. Rev. D]
  {10.1103/PhysRevD.102.023509}, 102, 023509

\bibitem[\protect\citeauthoryear{Abbott et~al.}{Abbott
  et~al.}{2020b}]{DES_Y1_Clusters}
Abbott T. M.~C.,  et~al., 2020b, \mn@doi [Phys. Rev. D]
  {10.1103/PhysRevD.102.023509}, 102, 023509

\bibitem[\protect\citeauthoryear{Abbott et~al.}{Abbott
  et~al.}{2021}]{LIGO_H0_2021}
Abbott R.,  et~al., 2021

\bibitem[\protect\citeauthoryear{Abdalla et~al.}{Abdalla
  et~al.}{2022}]{SNOWMASS_tensions_2022}
Abdalla E.,  et~al., 2022, in {2022 Snowmass Summer Study}.  (\mn@eprint
  {arXiv} {2203.06142})

\bibitem[\protect\citeauthoryear{Ade et~al.}{Ade
  et~al.}{2016}]{Planck_2015_XIV}
Ade P. A.~R.,  et~al., 2016, \mn@doi [Astron. Astrophys.]
  {10.1051/0004-6361/201525814}, 594, A14

\bibitem[\protect\citeauthoryear{Adelberger, Heckel  \& Nelson}{Adelberger
  et~al.}{2003}]{Adelberger_etal_2003}
Adelberger E.~G.,  Heckel B.~R.,   Nelson A.~E.,  2003, \mn@doi [Ann. Rev.
  Nucl. Part. Sci.] {10.1146/annurev.nucl.53.041002.110503}, 53, 77

\bibitem[\protect\citeauthoryear{Aiola et~al.}{Aiola
  et~al.}{2020}]{ACT_DR4_2020}
Aiola S.,  et~al., 2020, \mn@doi [JCAP] {10.1088/1475-7516/2020/12/047}, 12,
  047

\bibitem[\protect\citeauthoryear{Amendola}{Amendola}{2000}]{Amendola_2000}
Amendola L.,  2000, \mn@doi [Phys. Rev.] {10.1103/PhysRevD.62.043511}, D62,
  043511

\bibitem[\protect\citeauthoryear{Amendola}{Amendola}{2004}]{Amendola_2004}
Amendola L.,  2004, \mn@doi [Phys. Rev.] {10.1103/PhysRevD.69.103524}, D69,
  103524

\bibitem[\protect\citeauthoryear{Amendola \& Quercellini}{Amendola \&
  Quercellini}{2003}]{Amendola_Quercellini_2003}
Amendola L.,  Quercellini C.,  2003, \mn@doi [Phys. Rev.]
  {10.1103/PhysRevD.68.023514}, D68, 023514

\bibitem[\protect\citeauthoryear{Amendola \& Tocchini-Valentini}{Amendola \&
  Tocchini-Valentini}{2002}]{Amendola_Tocchini-Valentini_2002}
Amendola L.,  Tocchini-Valentini D.,  2002, in {37th Rencontres de Moriond on
  the Cosmological Model}. pp 407--410 (\mn@eprint {arXiv} {astro-ph/0205545})

\bibitem[\protect\citeauthoryear{{Amendola}, {Baldi}  \&
  {Wetterich}}{{Amendola} et~al.}{2008}]{Amendola_Baldi_Wetterich_2008}
{Amendola} L.,  {Baldi} M.,   {Wetterich} C.,  2008, \mn@doi [\prd]
  {10.1103/PhysRevD.78.023015}, \href
  {http://adsabs.harvard.edu/abs/2008PhRvD..78b3015A} {78, 023015}

\bibitem[\protect\citeauthoryear{Amendola, Barreiro  \& Nunes}{Amendola
  et~al.}{2014}]{Amendola_Barreiro_Nunes_2014}
Amendola L.,  Barreiro T.,   Nunes N.~J.,  2014, \mn@doi [Phys. Rev. D]
  {10.1103/PhysRevD.90.083508}, 90, 083508

\bibitem[\protect\citeauthoryear{Amon et~al.}{Amon et~al.}{2022}]{DES_Y3_Amon}
Amon A.,  et~al., 2022, \mn@doi [Phys. Rev. D] {10.1103/PhysRevD.105.023514},
  105, 023514

\bibitem[\protect\citeauthoryear{Archidiacono, Castorina, Redigolo  \&
  Salvioni}{Archidiacono et~al.}{2022}]{Archidiacono_etal_2022}
Archidiacono M.,  Castorina E.,  Redigolo D.,   Salvioni E.,  2022

\bibitem[\protect\citeauthoryear{Armendariz-Picon, Mukhanov  \&
  Steinhardt}{Armendariz-Picon et~al.}{2001}]{kessence}
Armendariz-Picon C.,  Mukhanov V.~F.,   Steinhardt P.~J.,  2001, \mn@doi [Phys.
  Rev.] {10.1103/PhysRevD.63.103510}, D63, 103510

\bibitem[\protect\citeauthoryear{Asgari et~al.}{Asgari
  et~al.}{2021}]{KiDS-1000}
Asgari M.,  et~al., 2021, \mn@doi [Astron. Astrophys.]
  {10.1051/0004-6361/202039070}, 645, A104

\bibitem[\protect\citeauthoryear{Baccigalupi, Matarrese  \&
  Perrotta}{Baccigalupi et~al.}{2000}]{Baccigalupi_Matarrese_Perrotta_2000}
Baccigalupi C.,  Matarrese S.,   Perrotta F.,  2000, \mn@doi [Phys. Rev.]
  {10.1103/PhysRevD.62.123510}, D62, 123510

\bibitem[\protect\citeauthoryear{Bacon et~al.}{Bacon et~al.}{2020}]{SKA}
Bacon D.~J.,  et~al., 2020, \mn@doi [Publ. Astron. Soc. Austral.]
  {10.1017/pasa.2019.51}, 37, e007

\bibitem[\protect\citeauthoryear{{Baldi}}{{Baldi}}{2011a}]{Baldi_2011a}
{Baldi} M.,  2011a, \mn@doi [\mnras] {10.1111/j.1365-2966.2010.17758.x}, \href
  {http://adsabs.harvard.edu/abs/2011MNRAS.411.1077B} {411, 1077}

\bibitem[\protect\citeauthoryear{{Baldi}}{{Baldi}}{2011b}]{Baldi_2011b}
{Baldi} M.,  2011b, \mn@doi [\mnras] {10.1111/j.1365-2966.2011.18263.x}, \href
  {http://adsabs.harvard.edu/abs/2011MNRAS.414..116B} {414, 116}

\bibitem[\protect\citeauthoryear{{Baldi}}{{Baldi}}{2012a}]{Baldi_2011c}
{Baldi} M.,  2012a, \mn@doi [\mnras] {10.1111/j.1365-2966.2011.20048.x}, \href
  {http://adsabs.harvard.edu/abs/2012MNRAS.420..430B} {420, 430}

\bibitem[\protect\citeauthoryear{{Baldi}}{{Baldi}}{2012b}]{CoDECS}
{Baldi} M.,  2012b, \mn@doi [\mnras] {10.1111/j.1365-2966.2012.20675.x}, \href
  {http://adsabs.harvard.edu/abs/2012MNRAS.422.1028B} {422, 1028}

\bibitem[\protect\citeauthoryear{{Baldi}}{{Baldi}}{2012c}]{Baldi_2012a}
{Baldi} M.,  2012c, \mn@doi [Annalen der Physik] {10.1002/andp.201200073},
  \href {http://adsabs.harvard.edu/abs/2012AnP...524..602B} {524, 602}

\bibitem[\protect\citeauthoryear{Baldi \& Simpson}{Baldi \&
  Simpson}{2017}]{Baldi_Simpson_2017}
Baldi M.,  Simpson F.,  2017, \mn@doi [Mon. Not. Roy. Astron. Soc.]
  {10.1093/mnras/stw2702}, 465, 653

\bibitem[\protect\citeauthoryear{{Baldi}, {Pettorino}, {Robbers}  \&
  {Springel}}{{Baldi} et~al.}{2010}]{Baldi_etal_2010}
{Baldi} M.,  {Pettorino} V.,  {Robbers} G.,   {Springel} V.,  2010, \mn@doi
  [\mnras] {10.1111/j.1365-2966.2009.15987.x}, \href
  {http://adsabs.harvard.edu/abs/2010MNRAS.403.1684B} {403, 1684}

\bibitem[\protect\citeauthoryear{Balkenhol et~al.}{Balkenhol
  et~al.}{2021}]{SPT-3G_2021}
Balkenhol L.,  et~al., 2021, \mn@doi [Phys. Rev. D]
  {10.1103/PhysRevD.104.083509}, 104, 083509

\bibitem[\protect\citeauthoryear{Barros, Amendola, Barreiro  \& Nunes}{Barros
  et~al.}{2019}]{Barros_etal_2019}
Barros B.~J.,  Amendola L.,  Barreiro T.,   Nunes N.~J.,  2019, \mn@doi [JCAP]
  {10.1088/1475-7516/2019/01/007}, 01, 007

\bibitem[\protect\citeauthoryear{Barros, Barreiro  \& Nunes}{Barros
  et~al.}{2020}]{Barros_etal_2020}
Barros B.~J.,  Barreiro T.,   Nunes N.~J.,  2020, \mn@doi [Phys. Rev. D]
  {10.1103/PhysRevD.101.023502}, 101, 023502

\bibitem[\protect\citeauthoryear{Bertotti, Iess  \& Tortora}{Bertotti
  et~al.}{2003}]{Bertotti_Iess_Tortora_2003}
Bertotti B.,  Iess L.,   Tortora P.,  2003, \mn@doi [Nature]
  {10.1038/nature01997}, 425, 374

\bibitem[\protect\citeauthoryear{Bird, Cholis, Mu\~noz, Ali-Ha\"\i{}moud,
  Kamionkowski, Kovetz, Raccanelli  \& Riess}{Bird
  et~al.}{2016}]{Bird_etal_2016}
Bird S.,  Cholis I.,  Mu\~noz J.~B.,  Ali-Ha\"\i{}moud Y.,  Kamionkowski M.,
  Kovetz E.~D.,  Raccanelli A.,   Riess A.~G.,  2016, \mn@doi [Phys. Rev.
  Lett.] {10.1103/PhysRevLett.116.201301}, 116, 201301

\bibitem[\protect\citeauthoryear{Bocquet et~al.}{Bocquet et~al.}{2019}]{SPT-SZ}
Bocquet S.,  et~al., 2019, \mn@doi [Astrophys. J.] {10.3847/1538-4357/ab1f10},
  878, 55

\bibitem[\protect\citeauthoryear{Bonometto, Mainini  \& Macci{\`o}}{Bonometto
  et~al.}{2015}]{Bonometto_etal_2015}
Bonometto S.~A.,  Mainini R.,   Macci{\`o} A.~V.,  2015, \mn@doi [Mon. Not.
  Roy. Astron. Soc.] {10.1093/mnras/stv1621}, 453, 1002

\bibitem[\protect\citeauthoryear{Boyarsky, Ruchayskiy  \&
  Shaposhnikov}{Boyarsky et~al.}{2009}]{Boyarsky_Ruchayskiy_Shaposhnikov_2009}
Boyarsky A.,  Ruchayskiy O.,   Shaposhnikov M.,  2009, \mn@doi [Ann. Rev. Nucl.
  Part. Sci.] {10.1146/annurev.nucl.010909.083654}, 59, 191

\bibitem[\protect\citeauthoryear{Brax \& Martin}{Brax \&
  Martin}{1999}]{Brax_Martin_1999}
Brax P.,  Martin J.,  1999, \mn@doi [Phys. Lett.]
  {10.1016/S0370-2693(99)01209-5}, B468, 40

\bibitem[\protect\citeauthoryear{Buchmueller, Doglioni  \& Wang}{Buchmueller
  et~al.}{2017}]{Buchmueller_Doglioni_Wang_2017}
Buchmueller O.,  Doglioni C.,   Wang L.~T.,  2017, \mn@doi [Nature Phys.]
  {10.1038/nphys4054}, 13, 217

\bibitem[\protect\citeauthoryear{Cai, Padilla  \& Li}{Cai
  et~al.}{2015}]{Cai_Padilla_Li_2015}
Cai Y.-C.,  Padilla N.,   Li B.,  2015, \mn@doi [Mon. Not. Roy. Astron. Soc.]
  {10.1093/mnras/stv777}, 451, 1036

\bibitem[\protect\citeauthoryear{Caprini \& Tamanini}{Caprini \&
  Tamanini}{2016}]{Caprini_Tamanini_2016}
Caprini C.,  Tamanini N.,  2016, \mn@doi [JCAP]
  {10.1088/1475-7516/2016/10/006}, 10, 006

\bibitem[\protect\citeauthoryear{Carr, Kuhnel  \& Sandstad}{Carr
  et~al.}{2016}]{Carr_Kuhnel_Sandstad_2016}
Carr B.,  Kuhnel F.,   Sandstad M.,  2016, \mn@doi [Phys. Rev. D]
  {10.1103/PhysRevD.94.083504}, 94, 083504

\bibitem[\protect\citeauthoryear{Chen, Vlah  \& White}{Chen
  et~al.}{2022}]{Chen_etal_2022}
Chen S.-F.,  Vlah Z.,   White M.,  2022, \mn@doi [JCAP]
  {10.1088/1475-7516/2022/02/008}, 02, 008

\bibitem[\protect\citeauthoryear{Contarini, Ronconi, Marulli, Moscardini,
  Veropalumbo  \& Baldi}{Contarini et~al.}{2019}]{Contarini_etal_2019}
Contarini S.,  Ronconi T.,  Marulli F.,  Moscardini L.,  Veropalumbo A.,
  Baldi M.,  2019, \mn@doi [Mon. Not. Roy. Astron. Soc.]
  {10.1093/mnras/stz1989}, 488, 3526

\bibitem[\protect\citeauthoryear{Contarini, Marulli, Moscardini, Veropalumbo,
  Giocoli  \& Baldi}{Contarini et~al.}{2020}]{Contarini_etal_2020}
Contarini S.,  Marulli F.,  Moscardini L.,  Veropalumbo A.,  Giocoli C.,
  Baldi M.,  2020, \mn@doi [Monthly Notices of the Royal Astronomical Society]
  {10.1093/mnras/stab1112}, 504, 5021

\bibitem[\protect\citeauthoryear{Copeland, Liddle  \& Wands}{Copeland
  et~al.}{1998}]{Copeland_etal_1998}
Copeland E.~J.,  Liddle A.~R.,   Wands D.,  1998, \mn@doi [Phys. Rev.]
  {10.1103/PhysRevD.57.4686}, D57, 4686

\bibitem[\protect\citeauthoryear{{Davis}, {Efstathiou}, {Frenk}  \&
  White}{{Davis} et~al.}{1985}]{Davis_etal_1985}
{Davis} M.,  {Efstathiou} G.,  {Frenk} C.~S.,   White S.~D.,  1985, \mn@doi
  [Astrophys.J.] {10.1086/163168}, 292, 371

\bibitem[\protect\citeauthoryear{{Farrar} \& {Peebles}}{{Farrar} \&
  {Peebles}}{2004}]{Farrar2004}
{Farrar} G.~R.,  {Peebles} P.~J.~E.,  2004, \mn@doi [\apj] {10.1086/381728},
  \href {http://adsabs.harvard.edu/abs/2004ApJ...604....1F} {604, 1}

\bibitem[\protect\citeauthoryear{Ferlito, Vagnozzi, Mota  \& Baldi}{Ferlito
  et~al.}{2022}]{Ferlito_etal_2022}
Ferlito F.,  Vagnozzi S.,  Mota D.~F.,   Baldi M.,  2022, \mn@doi [Mon. Not.
  Roy. Astron. Soc.] {10.1093/mnras/stac649}, 512, 1885

\bibitem[\protect\citeauthoryear{Ferreira \& Joyce}{Ferreira \&
  Joyce}{1998}]{Ferreira_Joyce_1998}
Ferreira P.~G.,  Joyce M.,  1998, \mn@doi [Phys. Rev.]
  {10.1103/PhysRevD.58.023503}, D58, 023503

\bibitem[\protect\citeauthoryear{Garaldi, Baldi  \& Moscardini}{Garaldi
  et~al.}{2016}]{Garaldi_Baldi_Moscardini_2016}
Garaldi E.,  Baldi M.,   Moscardini L.,  2016, \mn@doi [JCAP]
  {10.1088/1475-7516/2016/01/050}, 01, 050

\bibitem[\protect\citeauthoryear{Gaskins}{Gaskins}{2016}]{Gaskins_2016}
Gaskins J.~M.,  2016, \mn@doi [Contemp. Phys.] {10.1080/00107514.2016.1175160},
  57, 496

\bibitem[\protect\citeauthoryear{Giocoli, Metcalf, Baldi, Meneghetti,
  Moscardini  \& Petkova}{Giocoli et~al.}{2015}]{Giocoli_etal_2015}
Giocoli C.,  Metcalf R.~B.,  Baldi M.,  Meneghetti M.,  Moscardini L.,
  Petkova M.,  2015, \mn@doi [Mon. Not. Roy. Astron. Soc.]
  {10.1093/mnras/stv1473}, 452, 2757

\bibitem[\protect\citeauthoryear{G\'omez-Valent, Pettorino  \&
  Amendola}{G\'omez-Valent et~al.}{2020}]{Gomez-Valent_Pettorino_Amendola_2020}
G\'omez-Valent A.,  Pettorino V.,   Amendola L.,  2020, \mn@doi [Phys. Rev. D]
  {10.1103/PhysRevD.101.123513}, 101, 123513

\bibitem[\protect\citeauthoryear{Green \& Kavanagh}{Green \&
  Kavanagh}{2021}]{Green_Kavanagh_2021}
Green A.~M.,  Kavanagh B.~J.,  2021, \mn@doi [J. Phys. G]
  {10.1088/1361-6471/abc534}, 48, 043001

\bibitem[\protect\citeauthoryear{Hahn \& Abel}{Hahn \& Abel}{2011}]{Music}
Hahn O.,  Abel T.,  2011, \mn@doi [Mon. Not. Roy. Astron. Soc.]
  {10.1111/j.1365-2966.2011.18820.x}, 415, 2101

\bibitem[\protect\citeauthoryear{Hamaus, Pisani, Choi, Lavaux, Wandelt  \&
  Weller}{Hamaus et~al.}{2020}]{Hamaus_etal_2020}
Hamaus N.,  Pisani A.,  Choi J.-A.,  Lavaux G.,  Wandelt B.~D.,   Weller J.,
  2020, \mn@doi [JCAP] {10.1088/1475-7516/2020/12/023}, 12, 023

\bibitem[\protect\citeauthoryear{Hamaus et~al.}{Hamaus
  et~al.}{2021}]{Hamaus_etal_2021}
Hamaus N.,  et~al., 2021, \mn@doi [arXiv: 2108.10347]
  {10.1051/0004-6361/202142073}

\bibitem[\protect\citeauthoryear{{Heymans} et~al.,}{{Heymans}
  et~al.}{2021}]{Heymans_etal_2021}
{Heymans} C.,  et~al., 2021, \mn@doi [\aap] {10.1051/0004-6361/202039063},
  \href {https://ui.adsabs.harvard.edu/abs/2021A&A...646A.140H} {646, A140}

\bibitem[\protect\citeauthoryear{{Hildebrandt} et~al.,}{{Hildebrandt}
  et~al.}{2017}]{Hildebrandt_etal_2017}
{Hildebrandt} H.,  et~al., 2017, \mn@doi [\mnras] {10.1093/mnras/stw2805},
  \href {http://adsabs.harvard.edu/abs/2017MNRAS.465.1454H} {465, 1454}

\bibitem[\protect\citeauthoryear{{Hildebrandt} et~al.,}{{Hildebrandt}
  et~al.}{2020}]{Hildebrandt_etal_2020}
{Hildebrandt} H.,  et~al., 2020, \mn@doi [\aap] {10.1051/0004-6361/201834878},
  \href {https://ui.adsabs.harvard.edu/abs/2020A&A...633A..69H} {633, A69}

\bibitem[\protect\citeauthoryear{Hu \& Sawicki}{Hu \&
  Sawicki}{2007}]{Hu_Sawicki_2007}
Hu W.,  Sawicki I.,  2007, \mn@doi [Phys. Rev.] {10.1103/PhysRevD.76.064004},
  D76, 064004

\bibitem[\protect\citeauthoryear{Hu, Barkana  \& Gruzinov}{Hu
  et~al.}{2000}]{Hu_Barkana_Gruzinov_2000}
Hu W.,  Barkana R.,   Gruzinov A.,  2000, \mn@doi [Phys. Rev. Lett.]
  {10.1103/PhysRevLett.85.1158}, 85, 1158

\bibitem[\protect\citeauthoryear{Huey \& Wandelt}{Huey \&
  Wandelt}{2006}]{Huey_Wandelt_2006}
Huey G.,  Wandelt B.~D.,  2006, \mn@doi [Phys. Rev.]
  {10.1103/PhysRevD.74.023519}, D74, 023519

\bibitem[\protect\citeauthoryear{Hui, Ostriker, Tremaine  \& Witten}{Hui
  et~al.}{2017}]{Hui_etal_2017}
Hui L.,  Ostriker J.~P.,  Tremaine S.,   Witten E.,  2017, \mn@doi [Phys. Rev.
  D] {10.1103/PhysRevD.95.043541}, 95, 043541

\bibitem[\protect\citeauthoryear{Ivezic et~al.}{Ivezic et~al.}{2008}]{LSST}
Ivezic Z.,  et~al., 2008, arXiv:0805.2366

\bibitem[\protect\citeauthoryear{{Joudaki} et~al.,}{{Joudaki}
  et~al.}{2017}]{Joudaki_etal_2017}
{Joudaki} S.,  et~al., 2017, \mn@doi [\mnras] {10.1093/mnras/stx998}, \href
  {https://ui.adsabs.harvard.edu/abs/2017MNRAS.471.1259J} {471, 1259}

\bibitem[\protect\citeauthoryear{{Joudaki} et~al.,}{{Joudaki}
  et~al.}{2018}]{Joudaki_etal_2018}
{Joudaki} S.,  et~al., 2018, \mn@doi [\mnras] {10.1093/mnras/stx2820}, \href
  {https://ui.adsabs.harvard.edu/abs/2018MNRAS.474.4894J} {474, 4894}

\bibitem[\protect\citeauthoryear{Kazantzidis \& Perivolaropoulos}{Kazantzidis
  \& Perivolaropoulos}{2018}]{Kazantzidis_Perivolaropoulos_2018}
Kazantzidis L.,  Perivolaropoulos L.,  2018, \mn@doi [Phys. Rev. D]
  {10.1103/PhysRevD.97.103503}, 97, 103503

\bibitem[\protect\citeauthoryear{Kesden \& Kamionkowski}{Kesden \&
  Kamionkowski}{2006}]{Kesden_Kamionkowski_2006}
Kesden M.,  Kamionkowski M.,  2006, \mn@doi [Phys. Rev.]
  {10.1103/PhysRevD.74.083007}, D74, 083007

\bibitem[\protect\citeauthoryear{Keselman, Nusser  \& Peebles}{Keselman
  et~al.}{2009}]{Keselman_Nusser_Peebles_2009}
Keselman J.~A.,  Nusser A.,   Peebles P. J.~E.,  2009, \mn@doi [Phys. Rev.]
  {10.1103/PhysRevD.80.063517}, D80, 063517

\bibitem[\protect\citeauthoryear{Keselman, Nusser  \& Peebles}{Keselman
  et~al.}{2010}]{Keselman_Nusser_Peebles_2010}
Keselman J.~A.,  Nusser A.,   Peebles P.,  2010, \mn@doi [Phys.Rev.]
  {10.1103/PhysRevD.81.063521}, D81, 063521

\bibitem[\protect\citeauthoryear{Laureijs et~al.}{Laureijs
  et~al.}{2011}]{Laureijs_etal_2011}
Laureijs R.,  et~al., 2011

\bibitem[\protect\citeauthoryear{Lesci et~al.}{Lesci
  et~al.}{2022}]{Lesci_etal_2022}
Lesci G.~F.,  et~al., 2022, \mn@doi [Astron. Astrophys.]
  {10.1051/0004-6361/202040194}, 659, A88

\bibitem[\protect\citeauthoryear{Lewis, Challinor  \& Lasenby}{Lewis
  et~al.}{2000}]{camb}
Lewis A.,  Challinor A.,   Lasenby A.,  2000, \mn@doi [Astrophys. J.]
  {10.1086/309179}, 538, 473

\bibitem[\protect\citeauthoryear{Li \& Barrow}{Li \&
  Barrow}{2010}]{Li_Barrow_2010b}
Li B.,  Barrow J.~D.,  2010, arXiv:1010.3748

\bibitem[\protect\citeauthoryear{Li \& Zhao}{Li \& Zhao}{2009}]{Li_Zhao_2009}
Li B.,  Zhao H.,  2009, \mn@doi [Phys. Rev.] {10.1103/PhysRevD.80.044027}, D80,
  044027

\bibitem[\protect\citeauthoryear{Lucchin \& Matarrese}{Lucchin \&
  Matarrese}{1985}]{Lucchin_Matarrese_1984}
Lucchin F.,  Matarrese S.,  1985, \mn@doi [Phys. Rev.]
  {10.1103/PhysRevD.32.1316}, D32, 1316

\bibitem[\protect\citeauthoryear{Macaulay, Wehus  \& Eriksen}{Macaulay
  et~al.}{2013}]{Macaulay_etal_2013}
Macaulay E.,  Wehus I.~K.,   Eriksen H.~K.,  2013, \mn@doi [Phys. Rev. Lett.]
  {10.1103/PhysRevLett.111.161301}, 111, 161301

\bibitem[\protect\citeauthoryear{Macci{\`o}, Mainini, Penzo  \&
  Bonometto}{Macci{\`o} et~al.}{2015}]{Maccio_etal_2015}
Macci{\`o} A.~V.,  Mainini R.,  Penzo C.,   Bonometto S.~A.,  2015, \mn@doi
  [Mon. Not. Roy. Astron. Soc.] {10.1093/mnras/stv1680}, 453, 1371

\bibitem[\protect\citeauthoryear{Mainini \& Bonometto}{Mainini \&
  Bonometto}{2004}]{Mainini_Bonometto_2004}
Mainini R.,  Bonometto S.~A.,  2004, \mn@doi [Phys. Rev. Lett.]
  {10.1103/PhysRevLett.93.121301}, 93, 121301

\bibitem[\protect\citeauthoryear{Marrod\'an~Undagoitia \&
  Rauch}{Marrod\'an~Undagoitia \& Rauch}{2016}]{MarrodanUndagoitia_Rauch_2016}
Marrod\'an~Undagoitia T.,  Rauch L.,  2016, \mn@doi [J. Phys. G]
  {10.1088/0954-3899/43/1/013001}, 43, 013001

\bibitem[\protect\citeauthoryear{Marulli, Veropalumbo, Garc\'\i{}a-Farieta,
  Moresco, Moscardini  \& Cimatti}{Marulli et~al.}{2021}]{Marulli_etal_2021}
Marulli F.,  Veropalumbo A.,  Garc\'\i{}a-Farieta J.~E.,  Moresco M.,
  Moscardini L.,   Cimatti A.,  2021, \mn@doi [Astrophys. J.]
  {10.3847/1538-4357/ac0e8c}, 920, 13

\bibitem[\protect\citeauthoryear{Moresco et~al.}{Moresco
  et~al.}{2022}]{Moresco_etal_2022}
Moresco M.,  et~al., 2022

\bibitem[\protect\citeauthoryear{Neyrinck}{Neyrinck}{2008}]{Neyrinck_2008}
Neyrinck M.~C.,  2008, \mn@doi [Mon. Not. Roy. Astron. Soc.]
  {10.1111/j.1365-2966.2008.13180.x}, 386, 2101

\bibitem[\protect\citeauthoryear{Nusser, Gubser  \& Peebles}{Nusser
  et~al.}{2005}]{Nusser_Gubser_Peebles_2005}
Nusser A.,  Gubser S.~S.,   Peebles P. J.~E.,  2005, \mn@doi [Phys. Rev.]
  {10.1103/PhysRevD.71.083505}, D71, 083505

\bibitem[\protect\citeauthoryear{{Pace}, {Baldi}, {Moscardini}, {Bacon}  \&
  {Crittenden}}{{Pace} et~al.}{2015}]{Pace_etal_2015}
{Pace} F.,  {Baldi} M.,  {Moscardini} L.,  {Bacon} D.,   {Crittenden} R.,
  2015, \mn@doi [\mnras] {10.1093/mnras/stu2513}, \href
  {http://adsabs.harvard.edu/abs/2015MNRAS.447..858P} {447, 858}

\bibitem[\protect\citeauthoryear{Penzo, Macci{\`o}, Baldi, Casarini  \&
  O{\~n}orbe}{Penzo et~al.}{2015}]{Penzo_etal_2015}
Penzo C.,  Macci{\`o} A.~V.,  Baldi M.,  Casarini L.,   O{\~n}orbe J.,  2015

\bibitem[\protect\citeauthoryear{Pettorino}{Pettorino}{2013}]{Pettorino_2013}
Pettorino V.,  2013, Phys. Rev. D 88,, 063519

\bibitem[\protect\citeauthoryear{Pettorino \& Baccigalupi}{Pettorino \&
  Baccigalupi}{2008}]{Pettorino_Baccigalupi_2008}
Pettorino V.,  Baccigalupi C.,  2008, \mn@doi [Phys. Rev.]
  {10.1103/PhysRevD.77.103003}, D77, 103003

\bibitem[\protect\citeauthoryear{Pettorino, Amendola, Baccigalupi  \&
  Quercellini}{Pettorino et~al.}{2012}]{Pettorino_etal_2012}
Pettorino V.,  Amendola L.,  Baccigalupi C.,   Quercellini C.,  2012, \mn@doi
  [Phys.Rev.] {10.1103/PhysRevD.86.103507}, D86, 103507

\bibitem[\protect\citeauthoryear{Pettorino, Amendola  \& Wetterich}{Pettorino
  et~al.}{2013}]{Pettorino_etal_2013}
Pettorino V.,  Amendola L.,   Wetterich C.,  2013, \mn@doi [Phys. Rev.]
  {10.1103/PhysRevD.87.083009}, D87, 083009

\bibitem[\protect\citeauthoryear{{Pisani}, {Sutter}, {Hamaus}, {Alizadeh},
  {Biswas}, {Wandelt}  \& {Hirata}}{{Pisani} et~al.}{2015}]{Pisani_etal_2015}
{Pisani} A.,  {Sutter} P.~M.,  {Hamaus} N.,  {Alizadeh} E.,  {Biswas} R.,
  {Wandelt} B.~D.,   {Hirata} C.~M.,  2015, preprint, \href
  {http://adsabs.harvard.edu/abs/2015arXiv150307690P} {} (\mn@eprint {arXiv}
  {1503.07690})

\bibitem[\protect\citeauthoryear{{Planck Collaboration} et~al.,}{{Planck
  Collaboration} et~al.}{2018}]{Planck_2018_VI}
{Planck Collaboration} et~al., 2018, ArXiv e-prints 1807.06209, \href
  {http://adsabs.harvard.edu/abs/2018arXiv180706209P} {}

\bibitem[\protect\citeauthoryear{Platen, van~de Weygaert  \& Jones}{Platen
  et~al.}{2007}]{Platen_etal_2007}
Platen E.,  van~de Weygaert R.,   Jones B. J.~T.,  2007, \mn@doi [Mon. Not.
  Roy. Astron. Soc.] {10.1111/j.1365-2966.2007.12125.x}, 380, 551

\bibitem[\protect\citeauthoryear{{Pollina}, {Baldi}, {Marulli}  \&
  {Moscardini}}{{Pollina} et~al.}{2016}]{Pollina_etal_2016}
{Pollina} G.,  {Baldi} M.,  {Marulli} F.,   {Moscardini} L.,  2016, \mn@doi
  [\mnras] {10.1093/mnras/stv2503}, \href
  {http://adsabs.harvard.edu/abs/2016MNRAS.455.3075P} {455, 3075}

\bibitem[\protect\citeauthoryear{Pourtsidou \& Tram}{Pourtsidou \&
  Tram}{2016}]{Pourtsidou_Tram_2016}
Pourtsidou A.,  Tram T.,  2016, \mn@doi [Phys. Rev. D]
  {10.1103/PhysRevD.94.043518}, 94, 043518

\bibitem[\protect\citeauthoryear{Pourtsidou, Skordis  \& Copeland}{Pourtsidou
  et~al.}{2013}]{Pourtsidou_etal_2013}
Pourtsidou A.,  Skordis C.,   Copeland E.,  2013, \mn@doi [Phys.Rev.]
  {10.1103/PhysRevD.88.083505}, D88, 083505

\bibitem[\protect\citeauthoryear{Ratra \& Peebles}{Ratra \&
  Peebles}{1988}]{Ratra_Peebles_1988}
Ratra B.,  Peebles P. J.~E.,  1988, \mn@doi [Phys. Rev.]
  {10.1103/PhysRevD.37.3406}, D37, 3406

\bibitem[\protect\citeauthoryear{Riess et~al.}{Riess
  et~al.}{2007}]{Riess_etal_2006}
Riess A.~G.,  et~al., 2007, \mn@doi [Astrophys. J.] {10.1086/510378}, 659, 98

\bibitem[\protect\citeauthoryear{Riess et~al.}{Riess et~al.}{2021}]{SHOES_2021}
Riess A.~G.,  et~al., 2021

\bibitem[\protect\citeauthoryear{Ronconi, Contarini, Marulli, Baldi  \&
  Moscardini}{Ronconi et~al.}{2019}]{Ronconi_etal_2019}
Ronconi T.,  Contarini S.,  Marulli F.,  Baldi M.,   Moscardini L.,  2019,
  \mn@doi [Mon. Not. Roy. Astron. Soc.] {10.1093/mnras/stz2115}, 488, 5075

\bibitem[\protect\citeauthoryear{Sahni}{Sahni}{2002}]{Sahni_2002}
Sahni V.,  2002, \mn@doi [Class. Quant. Grav.] {10.1088/0264-9381/19/13/304},
  19, 3435

\bibitem[\protect\citeauthoryear{Schuster, Hamaus, Pisani, Carbone, Kreisch,
  Pollina  \& Weller}{Schuster et~al.}{2019}]{Schuster_etal_2019}
Schuster N.,  Hamaus N.,  Pisani A.,  Carbone C.,  Kreisch C.~D.,  Pollina G.,
   Weller J.,  2019, \mn@doi [JCAP] {10.1088/1475-7516/2019/12/055}, 12, 055

\bibitem[\protect\citeauthoryear{Secco et~al.}{Secco
  et~al.}{2022}]{DES_Y3_Secco}
Secco L.~F.,  et~al., 2022, \mn@doi [Phys. Rev. D]
  {10.1103/PhysRevD.105.023515}, 105, 023515

\bibitem[\protect\citeauthoryear{Simpson}{Simpson}{2010}]{Simpson_2010}
Simpson F.,  2010, \mn@doi [Phys. Rev.] {10.1103/PhysRevD.82.083505}, D82,
  083505

\bibitem[\protect\citeauthoryear{Skordis, Pourtsidou  \& Copeland}{Skordis
  et~al.}{2015}]{Skordis_Pourtsidou_Copeland_2015}
Skordis C.,  Pourtsidou A.,   Copeland E.~J.,  2015, \mn@doi [Phys. Rev.]
  {10.1103/PhysRevD.91.083537}, D91, 083537

\bibitem[\protect\citeauthoryear{Smith et~al.}{Smith
  et~al.}{2003}]{Smith_etal_2003}
Smith R.~E.,  et~al., 2003, \mn@doi [Mon. Not. Roy. Astron. Soc.]
  {10.1046/j.1365-8711.2003.06503.x}, 341, 1311

\bibitem[\protect\citeauthoryear{Sotiriou \& Faraoni}{Sotiriou \&
  Faraoni}{2010}]{Sotiriou_Faraoni_2010}
Sotiriou T.~P.,  Faraoni V.,  2010, \mn@doi [Rev.Mod.Phys.]
  {10.1103/RevModPhys.82.451}, 82, 451

\bibitem[\protect\citeauthoryear{Springel}{Springel}{2005}]{gadget-2}
Springel V.,  2005, Mon. Not. Roy. Astron. Soc., 364, 1105

\bibitem[\protect\citeauthoryear{{Springel}, {White}, {Tormen}  \&
  {Kauffmann}}{{Springel} et~al.}{2001}]{Springel_etal_2001}
{Springel} V.,  {White} S.~D.~M.,  {Tormen} G.,   {Kauffmann} G.,  2001,
  \mn@doi [\mnras] {10.1046/j.1365-8711.2001.04912.x}, \href
  {http://adsabs.harvard.edu/abs/2001MNRAS.328..726S} {328, 726}

\bibitem[\protect\citeauthoryear{{Springel} et~al.,}{{Springel}
  et~al.}{2008}]{Aquarius}
{Springel} V.,  et~al., 2008, \mn@doi [\mnras]
  {10.1111/j.1365-2966.2008.14066.x}, \href
  {http://adsabs.harvard.edu/abs/2008MNRAS.391.1685S} {391, 1685}

\bibitem[\protect\citeauthoryear{{Sutter} et~al.,}{{Sutter}
  et~al.}{2015a}]{VIDE}
{Sutter} P.~M.,  et~al., 2015a, \mn@doi [Astronomy and Computing]
  {10.1016/j.ascom.2014.10.002}, \href
  {http://adsabs.harvard.edu/abs/2015A%26C.....9....1S} {9, 1}

\bibitem[\protect\citeauthoryear{{Sutter}, {Carlesi}, {Wandelt}  \&
  {Knebe}}{{Sutter} et~al.}{2015b}]{Sutter_etal_2015}
{Sutter} P.~M.,  {Carlesi} E.,  {Wandelt} B.~D.,   {Knebe} A.,  2015b, \mn@doi
  [\mnras] {10.1093/mnrasl/slu155}, \href
  {http://adsabs.harvard.edu/abs/2015MNRAS.446L...1S} {446, L1}

\bibitem[\protect\citeauthoryear{Takahashi, Sato, Nishimichi, Taruya  \&
  Oguri}{Takahashi et~al.}{2012}]{Takahashi_etal_2012}
Takahashi R.,  Sato M.,  Nishimichi T.,  Taruya A.,   Oguri M.,  2012, \mn@doi
  [Astrophys. J.] {10.1088/0004-637X/761/2/152}, 761, 152

\bibitem[\protect\citeauthoryear{{Tarrant}, {van de Bruck}, {Copeland}  \&
  {Green}}{{Tarrant} et~al.}{2012}]{Tarrant_etal_2012}
{Tarrant} E.~R.~M.,  {van de Bruck} C.,  {Copeland} E.~J.,   {Green} A.~M.,
  2012, \mn@doi [\prd] {10.1103/PhysRevD.85.023503}, \href
  {http://adsabs.harvard.edu/abs/2012PhRvD..85b3503T} {85, 023503}

\bibitem[\protect\citeauthoryear{Tinker, Robertson, Kravtsov, Klypin, Warren,
  Yepes  \& Gottlober}{Tinker et~al.}{2010}]{Tinker_etal_2010}
Tinker J.~L.,  Robertson B.~E.,  Kravtsov A.~V.,  Klypin A.,  Warren M.~S.,
  Yepes G.,   Gottlober S.,  2010, \mn@doi [Astrophys. J.]
  {10.1088/0004-637X/724/2/878}, 724, 878

\bibitem[\protect\citeauthoryear{Tr\"oster et~al.}{Tr\"oster
  et~al.}{2020}]{Troster_etal_2020}
Tr\"oster T.,  et~al., 2020, \mn@doi [Astron. Astrophys.]
  {10.1051/0004-6361/201936772}, 633, L10

\bibitem[\protect\citeauthoryear{{Troxel} et~al.,}{{Troxel}
  et~al.}{2018}]{Troxel_etal_2018}
{Troxel} M.~A.,  et~al., 2018, \mn@doi [\prd] {10.1103/PhysRevD.98.043528},
  \href {https://ui.adsabs.harvard.edu/abs/2018PhRvD..98d3528T} {98, 043528}

\bibitem[\protect\citeauthoryear{Vagnozzi, Visinelli, Mena  \& Mota}{Vagnozzi
  et~al.}{2020}]{Vagnozzi_etal_2020}
Vagnozzi S.,  Visinelli L.,  Mena O.,   Mota D.~F.,  2020, \mn@doi [Mon. Not.
  Roy. Astron. Soc.] {10.1093/mnras/staa311}, 493, 1139

\bibitem[\protect\citeauthoryear{Verza, Pisani, Carbone, Hamaus  \&
  Guzzo}{Verza et~al.}{2019}]{Verza_etal_2019}
Verza G.,  Pisani A.,  Carbone C.,  Hamaus N.,   Guzzo L.,  2019, \mn@doi
  [JCAP] {10.1088/1475-7516/2019/12/040}, 12, 040

\bibitem[\protect\citeauthoryear{{Vikhlinin} et~al.,}{{Vikhlinin}
  et~al.}{2009}]{Vikhlinin_etal_2009}
{Vikhlinin} A.,  et~al., 2009, \mn@doi [\apj] {10.1088/0004-637X/692/2/1060},
  \href {http://adsabs.harvard.edu/abs/2009ApJ...692.1060V} {692, 1060}

\bibitem[\protect\citeauthoryear{Voivodic, Lima, Llinares  \& Mota}{Voivodic
  et~al.}{2017}]{Voivodic_etal_2017}
Voivodic R.,  Lima M.,  Llinares C.,   Mota D.~F.,  2017, \mn@doi [Phys. Rev.
  D] {10.1103/PhysRevD.95.024018}, 95, 024018

\bibitem[\protect\citeauthoryear{Weinberg}{Weinberg}{1989}]{Weinberg_1989}
Weinberg S.,  1989, \mn@doi [Rev. Mod. Phys.] {10.1103/RevModPhys.61.1}, 61, 1

\bibitem[\protect\citeauthoryear{Wetterich}{Wetterich}{1988}]{Wetterich_1988}
Wetterich C.,  1988, \mn@doi [Nucl. Phys.] {10.1016/0550-3213(88)90193-9},
  B302, 668

\bibitem[\protect\citeauthoryear{Wetterich}{Wetterich}{1995}]{Wetterich_1995}
Wetterich C.,  1995, Astron. Astrophys., 301, 321

\bibitem[\protect\citeauthoryear{Will}{Will}{2014}]{Will_2014}
Will C.~M.,  2014, \mn@doi [Living Rev. Rel.] {10.12942/lrr-2014-4}, 17, 4

\bibitem[\protect\citeauthoryear{Wittner, Laverda, Piattella  \&
  Amendola}{Wittner et~al.}{2020}]{Wittner_etal_2020}
Wittner M.,  Laverda G.,  Piattella O.~F.,   Amendola L.,  2020, \mn@doi [JCAP]
  {10.1088/1475-7516/2020/07/019}, 07, 019

\bibitem[\protect\citeauthoryear{Wong et~al.}{Wong
  et~al.}{2020}]{Wong_etal_2020}
Wong K.~C.,  et~al., 2020, \mn@doi [Mon. Not. Roy. Astron. Soc.]
  {10.1093/mnras/stz3094}, 498, 1420

\bibitem[\protect\citeauthoryear{van~de Bruck \& Thomas}{van~de Bruck \&
  Thomas}{2019}]{VanDeBruck_Thomas_2019}
van~de Bruck C.,  Thomas C.~C.,  2019, \mn@doi [Phys. Rev. D]
  {10.1103/PhysRevD.100.023515}, 100, 023515

\makeatother
\end{thebibliography}

\label{lastpage}

\end{document}